\documentstyle[aps,amsmath]{revtex}
 
\begin{document}
\draft

\title{
Polarization and localization in insulators: Generating function approach
}

\author{
Ivo Souza, Tim Wilkens, and Richard M. Martin}
\address{Department of Physics and Materials Research
Laboratory, University of Illinois, Urbana, Illinois 61801}
\date{\today}
\maketitle

\begin{abstract}

We develop the theory and practical expressions for the full quantum-mechanical
distribution of the intrinsic macroscopic polarization of an insulator in terms
of the ground state wavefunction. The central quantity is a cumulant generating
function which yields, upon successive differentiation, all the cumulants and 
moments of the probability distribution of the center
of mass ${\bf X}/N$ of the electrons, defined appropriately to remain valid for
extended systems obeying twisted boundary conditions.
The first moment is the average polarization, where we recover the well-known 
Berry phase expression. The second cumulant gives the mean-square fluctuation 
of the polarization, which defines an electronic localization length $\xi_i$
along each direction $i$:
$\xi_i^2 = {( \left< X_i^2 \right> - {\left< X_i \right>}^2)}/N$. It follows 
from the fluctuation-dissipation theorem that in the thermodynamic limit 
$\xi_i$ diverges for metals and is a finite, measurable quantity for 
insulators.  It is possible to 
define for insulators maximally-localized ``many-body Wannier functions'', 
which for large $N$ become localized in 
disconnected regions of the high-dimensional configuration space, establishing
a direct connection with Kohn's theory of the insulating state. 
Interestingly,
the expression for $\xi_i^2$, which involves the second derivative of the 
wavefunction with respect to the boundary conditions, is 
directly analogous to Kohn's formula for the ``Drude weight'' as the second
derivative of the energy.
\end{abstract}

\section{INTRODUCTION}\label{intro}

An insulator is distinguished from a conductor at zero temperature by 
its vanishing dc conductivity and 
its ability to sustain a
macroscopic polarization, both with and without an applied electric 
field\cite{landau_em,jackson}. In the classical theory of electromagnetism in
materials this distinction is often cast in terms of the difference
between
``free charges'' that carry the dc current in a conductor and  
polarizable
``bound charges'' in an insulator\cite{jackson,am}.
Such a description conflicts with
the fact that, even in highly ionic 
solids, the electrons are not well localized near the ions, and there is
appreciable
interpenetration between the ionic charge 
densities\cite{szigeti,harrison80,jennison,pendry}.
The inadequacy of such a textbook picture is particularly striking in the 
case of covalent insulators, whose charge density is
delocalized, as in metals\cite{am}. Therefore the qualitative 
difference between metals and insulators is not apparent from inspection of the
charge distribution, and the correct notion of electronic localization in 
insulators versus delocalization in conductors must be sought elsewhere.

As shown by Kohn\cite{kohn,kohn68}, localization 
is a property of the many-electron wavefunction: insulating behavior arises 
whenever the ground state wavefunction of an extended system breaks up into a 
sum of functions $\Psi_{\bf M}$ which are localized in essentially disconnected
regions ${\cal R}_{\bf M}$ of the high-dimensional configuration space. When
using periodic boundary conditions on a supercell containing $N$ electrons, an
insulating wavefunction can be written as

\begin{equation}
\label{psi_insulator}
\Psi({\bf x}_1,...,{\bf x}_N) = 
\sum_{{\bf M}=-{\boldsymbol \infty}}^{+{\boldsymbol \infty}} 
\Psi_{\bf M}({\bf x}_1,...,{\bf x}_N),
\end{equation}

\noindent where for a large supercell $\Psi_{\bf M}$ and $\Psi_{{\bf M}'}$ have
an exponentially small overlap for ${\bf M}' \not= {\bf M}$. Hence, electronic localization 
in insulators does not occur in real space (charge density) but in configuration 
space (wavefunction). Kohn argued that such disconnectedness is in fact the signature of an 
insulating wavefunction.

The other aspect of insulators, macroscopic polarization, is a subject about which there has 
been much debate. In a finite system with open boundary conditions, it is simply the dipole 
moment of the charge distribution divided by the volume\cite{landau_em,jackson}.
In the case of an extended system, such as a crystalline solid,
the situation is far less clear: although 
there exist well-known expressions for dielectric response 
functions\cite{vinay,pick}, the very definition of macroscopic 
polarization as
a bulk property independent of surface termination remained controversial for a
long time. Only in recent years has a theory of polarization 
emerged - the so-called Berry phase 
formulation\cite{ksv,vks,om94,resta94,resta_lect,resta98} - 
for the average bulk polarization in terms of the ground state
wavefunction of an insulating crystal. This theory shows that in general the 
information about the macroscopic polarization of a periodic system is not in 
the charge density, but in the wavefunction.
This important finding is consistent with the well-known fact 
that the dipole moment of a periodic, continuous charge distribution 
is ill-defined\cite{rmm74,hirst}, since the expressions for 
the first moment of the charge distribution are 
valid for finite systems but do not have a well-defined 
thermodynamic limit independent of the surface.
The only instance where a meaningful dipole moment can be assigned to a
unit cell of the crystal is when the charge distribution in the unit cell can 
be resolved into contributions which are localized
in non-overlapping regions and can be ascribed to identifiable atoms (ions,
molecules)\cite{am} (the so-called ``Clausius-Mossotti limit'').
However, such a limit is rather unrealistic for most insulators, with the
possible exception of some organic and molecular crystals.

In this paper we present a comprehensive theory of polarization and 
localization in insulators, which generalizes the Berry phase theory and merges
with Kohn's theory of localization in the insulating state.  By introducing
the generating function formalism, we show the rigorous connections between
polarization and localization, establish relations to experimentally
measureable quantities, and provide  formulas for practical calculations.

The paper is organized as follows: in Section \ref{recent_develop} we
summarize some of the recent developments in the field, pointing to some of the
open issues which will be addressed in this work, and collect 
relevant equations for later reference. In Section \ref{moment_gen_fnct} is 
introduced the central concept 
in the present formulation: the generating function applied to the 
quantum probability distribution of the macroscopic polarization.
The simple case of a finite system is discussed first, and then we show how to
generalize the expressions to deal with 
extended systems.
In Section \ref{fdt} we establish the fluctuation-dissipation relation
between the quadratic quantum fluctuations of the polarization and the
absorptive part of the conductivity for an extended system.
In Section \ref{fluct_ins_cond} we discuss the qualitative differences between
the quadratic fluctuations in insulators and in conductors in terms of a
properly defined localization length; for insulators we derive an inequality
involving the localization length and the optical gap.
In Section \ref{connection_kohn} we introduce ``many-body Wannier functions''
and establish the connection between the
present formalism and Kohn's theory of localization in the insulating state.
Discretized formulas for the polarization and localization which can be used
in numerical many-body calculations are derived in Section \ref{discrete}.
A discussion of our results is presented in Section \ref{discussion}.

\section{Overview of recent developments}
\label{recent_develop}

A meaningful definition of average macroscopic polarization which is generally
applicable - to both finite and extended systems - and relates closely to the 
way polarization is experimentally measured, can be obtained by taking as the 
more basic concept the {\it change} in polarization induced by a slow change in
some parameter $\lambda$ in the hamiltonian\cite{rmm72,resta92}. The resulting
expressions for an extended system are in terms of the derivative 
$\partial {\bf P}_{\rm bulk}^{(\lambda)} / \partial \lambda$, 
and Resta\cite{resta92} proposed to calculate the finite change in bulk
polarization as 

\begin{equation}
\label{delta_P}
\Delta {\bf P}_{\rm bulk} = 
\int_0^1 d \lambda \frac{\partial {\bf P}_{\rm bulk}^{(\lambda)}} 
{\partial \lambda}.
\end{equation}

\noindent Considering the variation in $\lambda$ as an adiabatic time 
evolution, $\partial {\bf P}^{(\lambda)}_{\rm bulk} / \partial \lambda$ is the
spatially-averaged adiabatic current flowing through the bulk; thus this 
equation gives $\Delta {\bf P}_{\rm bulk}$ as an integrated bulk current.
According to classical electrodynamics\cite{landau_em,jackson}, this 
adiabatic polarization current is

\begin{equation}
\label{current1}
\frac{{\partial} {\bf P}_{\rm bulk}^{(\lambda)}}{\partial \lambda} = 
\frac{1}{V} \int_V d {\bf r} {\bf j}_{\rm bulk}^{(\lambda)} =
\frac{1}{V} {\bf J}^{(\lambda)},
\end{equation}

\noindent where $V$ is the volume of the system and 
${\bf j}_{\rm bulk}^{(\lambda)}$ is the current density in the bulk,
where $\lambda$ plays the role of time in the usual expression for the current
(for a derivation see for example Ref. \cite{om94}). ${\bf J}^{(\lambda)}$
can be expressed as the adiabatic limit of a Kubo formula for the
current\cite{resta92,vogl78}. In its usual form the Kubo formula involves a 
summation over all eigenstates (ground state and excited states),
which not only makes it impractical for 
actual calculations, but also fails to reflect the fact that bulk polarization
is a ground state property. 

In the Berry phase theory of polarization\cite{ksv,vks,om94}, the 
Kubo formula is recast in a form which only depends on the ground state 
wavefunction; in order to arrive at such an expression, it is convenient to 
impose {\it twisted boundary conditions} over the volume $V$
on the many-body wavefunction, which becomes labelled by ${\bf k}$:

\begin{equation}
\label{twisted}
\Psi_{\rm \bf k}^{(\lambda)}({\bf x}_1,...,{\bf x}_i + {\bf L},...,{\bf x}_N) =
e^{\imath {\bf k} \cdot {\bf L}} 
\Psi_{\rm \bf k}^{(\lambda)}({\bf x}_1,...,{\bf x}_i,...,{\bf x}_N),
\end{equation}

\noindent where $N$ is the number of electrons in the system, 
${\bf L} = (L_1,0,0)$, for example, and
$- \pi / L_i \leq k_i \leq \pi / L_i$. 
It is convenient to introduce the following wavefunction:

\begin{equation}
\label{phi}
\left. \left| \Phi_{\bf k}^{(\lambda)} \right. \right> = 
e^{-\imath {\bf k} \cdot \hat{\bf X}} 
\left. \left| \Psi_{\bf k}^{(\lambda)} \right. \right>,
\end{equation}

\noindent where $\hat{\bf X} = \sum_{i=1}^N \hat{\bf x}_i$, so that
$\hat{\bf X}/N$ is the position operator for the center of mass of the $N$ 
electrons in the volume $V$. $\Phi_{\bf k}^{(\lambda)}$ can be regarded as the
many-body analogue of 
the cell-periodic part of the Bloch function in the single-electron case. It 
obeys periodic boundary conditions, and the ${\bf k}$-dependence is transfered 
from the boundary conditions to the hamiltonian: if 
$\Psi^{(\lambda)}_{\bf k}$ is the ground state of an hamiltonian 
$\hat{H}^{(\lambda)}$, then $\Phi_{\bf k}^{(\lambda)}$ is the ground 
state of the hamiltonian

\begin{equation}
\label{h_k}
\hat{H}^{(\lambda)}({\bf k}) = e^{-\imath {\bf k} \cdot \hat{\bf X}} 
\hat{H}^{(\lambda)} e^{\imath {\bf k} \cdot \hat{\bf X}}, 
\end{equation}

\noindent which, for a non-relativistic hamiltonian without spin-orbit 
coupling, can be obtained from $\hat{H}^{(\lambda)}$ by performing the
gauge transformation $\hat{\bf p}_i \rightarrow \hat{\bf p}_i + \hbar {\bf k}$
on the momentum operator of each particle
(for the single-particle analogue, see Ref. \cite{blount}.) 
Using the function $\Phi_{\bf k}^{(\lambda)}$, the Kubo formula for the 
adiabatic current ${\bf J}_{\bf k}^{(\lambda)}$ for a particular choice of 
twisted boundary conditions can be expressed 
as\cite{thouless83,niu84,resta_lect}

\begin{equation}
\label{adiabatic_current}
{\bf J}^{(\lambda)}_{\bf k} = 2 q_e {\rm Im}
\left< \left. {\partial}_{\bf k} \Phi_{\bf k}^{(\lambda)} \right| 
{\partial}_{\lambda} 
\Phi_{\bf k}^{(\lambda)} \right>,
\end{equation}

\noindent which indeed only depends on the ground state wavefunction. 
Substituting 
this expression into Eq. \ref{current1} and using Eq. \ref{delta_P}, it can be
shown that the net change in polarization along the path parametrized by 
$\lambda$ is 

\begin{equation}
\label{delta_P_integral}
{(\Delta P_{\rm el})}_i = \frac{\imath q_e}{{(2 \pi)}^3} \int d {\bf k}
\int_0^1 d \lambda \left[ \left< \left. \partial_{\lambda} 
\Phi_{\bf k}^{(\lambda)}
\right| \partial_{k_i} \Phi_{\bf k}^{(\lambda)} \right> -
\left< \left. \partial_{k_i} \Phi_{\bf k}^{(\lambda)}
\right| \partial_{\lambda} \Phi_{\bf k}^{(\lambda)} \right> \right],
\end{equation}

\noindent where the integral in ${\bf k}$ is over all twisted boundary 
conditions. 

As expected for a measurable quantity, the above expression, as well as 
Eq. \ref{adiabatic_current}, are invariant under gauge transformations of 
the form 

\begin{equation}
\label{gauge_trans}
\Psi_{\bf k}^{(\lambda)} \rightarrow e^{\imath \varphi({\bf k},\lambda)}
\Psi_{\bf k}^{(\lambda)},
\end{equation}

\noindent where $\varphi({\bf k},\lambda)$ is a smooth, real function.
The ground state wavefunctions $\Psi$ at ${\bf k}$ and 
${\bf k}+{\bf G}$, where ${\bf G}$ is a basis vector of the reciprocal lattice
of the cell of volume $V$,
can differ at most by a global phase factor, since the boundary conditions are
the same and the ground state is assumed to be non-degenerate\cite{niu84}:

\begin{equation}
\label{global_topology}
\Psi_{\bf k}^{(\lambda)} = e^{\imath \Theta({\bf k},\lambda; {\bf G})}
\Psi_{{\bf k}+{\bf G}}^{(\lambda)}.
\end{equation} 

\noindent If $\Theta({\bf k},\lambda; {\bf G})$, which is at our disposal, is 
chosen to be independent of $\lambda$, then it can be shown that the net change
in polarization becomes simply

\begin{equation}
\label{two_point}
\Delta {\bf P}_{\rm el} = {\bf P}_{\rm el}^{(1)} - {\bf P}_{\rm el}^{(0)},
\end{equation}

\noindent where

\begin{equation}
\label{delta_P_om94}
{\bf P}_{\rm el}^{(\lambda)} = 
\frac{\imath q_e}{{(2 \pi)}^3} \int d {\bf k} 
 \left< \left. \Phi_{\bf k}^{(\lambda)} \right| {\partial}_{\bf k} 
\Phi_{\bf k}^{(\lambda)} \right>,
\end{equation}

\noindent and again the integral in ${\bf k}$ is over all twisted boundary 
conditions. Eqs. \ref{delta_P_integral}, \ref{two_point}, and 
\ref{delta_P_om94}
form the central result of Ref. \cite{om94}, which gives the many-body
generalization of the Berry phase theory of polarization, originally 
formulated by King-Smith and Vanderbilt for independent 
electrons\cite{ksv,vks}. In the derivation it is assumed that the 
ground state is isolated from the excited states by a finite energy gap, and 
that there are no long-range correlations\cite{niu84}. In the 
independent-electron theory the Berry phase formula in terms of Bloch functions
can be recast in terms of localized Wannier functions\cite{wannier37}, yielding
the intuitive result that the electronic polarization is given by the sum of 
the centers of charge of the occupied Wannier functions\cite{ksv,vks}.
Unfortunately, Wannier functions are only defined in a one-electron framework. 
In Section \ref{loc_config_space} we will introduce functions which are in some
sense the many-body counterpart of Wannier functions (we will term them
``many-body Wannier functions''), and can be chosen to be localized in
configuration space. In terms of those functions, a localized description of
polarization can be presented for the correlated case. Moreover, they establish
the link between the Berry phase theory of polarization and Kohn's theory of 
localization\cite{kohn,kohn68}.

The following comments should be made about the above equations:
unlike Eq. \ref{delta_P_integral}, which involves an integral over $\lambda$, 
Eqs. \ref{two_point} and \ref{delta_P_om94} only depend on the endpoints,
$\lambda = 0$ and $\lambda = 1$. The tradeoff is that
whereas the former gives the exact change in polarization along the path,
the latter give it only modulo a ``quantum''\cite{ksv,vks,om94} (this 
``quantum of polarization'' is discussed in Appendix \ref{quantum}). A related
aspect is the behavior of the equations under gauge transformations: 
unlike Eqs. \ref{adiabatic_current} and \ref{delta_P_integral}, 
Eqs. \ref{two_point} and \ref{delta_P_om94}
are not completely gauge-invariant. As mentioned previously, they were obtained
by assuming $\partial_{\lambda} \Theta(\bf{k},\lambda;{\bf G})=0$, and 
therefore the resulting $\Delta {\bf P}_{\rm el}$
is only invariant (modulo
the quantum) under transformations which preserve the condition
$\Theta({\bf k},\lambda=1;{\bf G}) = \Theta({\bf k},\lambda=0;{\bf G})$.
Moreover, in order to be able to interpret Eq. \ref{delta_P_om94} at a 
{\it single} $\lambda$ 
as the electronic polarization\cite{vks,foot_one_point},
one has to impose the stronger condition 
$\Theta({\bf k},\lambda;{\bf G}) \equiv 0$, or 
$\Psi_{{\bf k}+{\bf G}}^{(\lambda)} = \Psi_{\bf k}^{(\lambda)}$. Gauges
that obey this condition are known as ``periodic gauges''\cite{ksv}. 

It has been pointed out by Aligia\cite{aligia99a} that this analysis needs to 
be modified in cases where there is a fractional number of electrons per 
primitive cell. The idea of a ``periodic gauge'' needs to be extended to relate
wavefunctions separated by multiples of the smallest ${\bf G}$, and the 
integral in Eq. \ref{delta_P_om94} needs to be extended accordingly.
For the sake of simplicity, in the main text we will assume integer filling, 
and in Appendix \ref{frac_fill} we indicate how to modify the formulas
in order to deal with fractional filling.

We note that in its present form the Berry phase theory of polarization only
gives the {\it average} of the quantum distribution of the macroscopic
polarization. In Sections \ref{gen_extended} and \ref{gi_cumulant} we will
extended it to deal with the full distribution. In particular, its quadratic 
spread will turn out to be a very relevant quantity,
since it is intimately related to electronic localization. To our
knowledge this was first discussed by Kudinov\cite{kudinov},
who proposed to measure the degree of localization in insulators precisely in 
terms of the mean-square quantum fluctuation of the ground state polarization. 
Kudinov proposed a criterion to differentiate between insulators and conductors
based on the scaling with sample volume $V$ of the mean square quantum 
fluctuation of the net dipole moment, 
$\left< {\Delta \hat{\bf d}}^2 \right> = 
\left< {\hat d}^2 \right> - {\left< \hat{\bf d} \right>}^2$;
here $\hat{\bf d}$ is the dipole moment operator,
$\left< ... \right>$ means the expectation value over the ground state,
and $\Delta \hat{\bf d} = \hat{\bf d} - \left< \hat{\bf d} \right>$.
Using the fluctuation-dissipation theorem\cite{callen,callen52,landau}, 
Kudinov related this quantity to the optical conductivity: 

\begin{equation}
\label{fdt1}
\frac {\left< {\Delta \hat{d_i}}^2 \right>}{V} =
\frac {\hbar}{\pi} \int_0^{\infty} d \omega {\rm Im} {\chi}_{ii}(\omega),
\end{equation}

\noindent where ${\chi}(\omega)$ is the electric susceptibility tensor and
$i=x,y,z$. Using Eq. \ref{fdt1}, Kudinov showed that as $V \rightarrow \infty$,
$\left< {\Delta \hat{\bf d}}^2 \right>/V$ remains finite in insulators, whereas
it diverges in conductors.
Only finite systems with open boundary conditions were considered, and thus the
issue of how to deal with the polarization in extended systems was not 
addressed. Our formulation leads to similar expressions, but with carefully 
defined bulk quantities which have a well-defined thermodynamic limit.

The ideas from the Berry phase theory of polarization have recently been
extended in order to address the problem of localization. This effort was
initiated in Ref. \cite{marzari97}, where non-interacting electron systems with
a band gap were considered. For such systems it is natural to attempt to 
quantify the degree of localization of the electrons in terms of the spread of
the occupied Wannier functions. Marzari and Vanderbilt proposed to measure that
spread via the quantity

\begin{equation}
\label{omega}
\Omega = \sum_{n=1}^M \left[ {\left< r^2 \right>}_n - 
         {\left< {\bf r} \right>}^2_n \right],
\end{equation}

\noindent where ${\left< ... \right>}_n$ means the expectation value over the
$n^{\rm th}$ occupied Wannier function in the unit cell (whose total number 
$M$ equals the number of filled bands.) Since the electronic polarization is 
given by the sum of the centers of charge of the occupied Wannier 
functions\cite{ksv,vks}, this expression is very appealing in its 
interpretation as the spread of the charge distribution of the Wannier 
functions. It should be noted however, that unlike the sum of the centers of 
charge, the sum of the quadratic spreads is not invariant under gauge 
transformations of the Wannier functions\cite{marzari97}, and so $\Omega$ 
cannot be used directly as a measure of any physical quantity. Nevertheless, 
Marzari and Vanderbilt were able to decompose it into a sum of two positive 
terms: a gauge-invariant part, $\Omega_{\rm I}$, plus a gauge-dependent term, 
which they minimized to obtain maximally-localized Wannier functions.

We will show that $\Omega_{\rm I}$ gives the mean-square fluctuation of the
bulk polarization, thus obeying a relation analogous to Eq. \ref{fdt1}
(see Appendix \ref{omega_I_fluct}). Moreover,
in the same way that $\Omega$ measures the spread of the Wannier functions, 
$\Omega_{\rm I}$ measures the spread of Kohn's functions 
$\Psi_{\bf M}$\cite{kohn,kohn68}, which can be interpreted as 
maximally-localized ``many-body Wannier functions''; this is discussed in 
Section \ref{loc_config_space} and Appendix \ref{omega_i_many_body}.

In Ref. \cite{marzari97} it was shown that $\Omega_{\rm I}$ be rewritten as

\begin{equation}
\label{omega_I}
\Omega_{\rm I} = \frac{v}{{(2 \pi)}^3} \int_{\rm BZ} d {\bf k} {\rm Tr} 
g({\bf k}),
\end{equation}

\noindent where Tr denotes the trace, $v$ is the volume of the unit cell, the 
integral is over the Brillouin zone, and $g({\bf k})$ is the tensor

\begin{equation}
\label{metric}
g_{ij}({\bf k}) = {\rm Re} \sum_{n=1}^M  \left< \left. \partial_{k_i} 
u_{n \bf k} \right|
\partial_{k_j} u_{m \bf k} \right> -
\sum_{n=1}^M \sum_{m=1}^M \left< \left. \partial_{k_i} u_{n \bf k} \right| 
u_{m \bf k} \right> \left< \left. u_{m \bf k} \right| \partial_{k_j} 
u_{n \bf k} \right>,
\end{equation} 

\noindent where $u_{n {\bf k}}$ is the cell-periodic part of the Bloch 
function. This tensor is a metric which can be used to 
determine the ``quantum distance'' along a given path in in 
${\bf k}$-space\cite{marzari97}.
In Section \ref{gi_cumulant} we generalize this tensor to the many-body case,
and in Section \ref{fdt} we relate it to the measurable polarization
fluctuations via the fluctuation-dissipation relation.

All of the above expressions for the polarization and localization involve
integrals over ${\bf k}$. 
More recently, alternative expressions have been proposed which use only
periodic boundary conditions 
(${\bf k}=0$)\cite{resta_lect,resta98,resta99}. These are sometimes called
``single-point'' formulas. The basic quantity in this formulation is, in one 
dimension (1D),

\begin{equation}
\label{z_N}
z_N = \left< \Psi_{{\bf k}=0} \left| e^{\imath (2 \pi / L) \hat{X}} \right| 
\Psi_{{\bf k}=0} \right>,
\end{equation}

\noindent where $\Psi_{{\bf k}=0}$ is the ground state many-body wavefunction 
obeying periodic boundary conditions over a cell of length $L$ with $N$ 
electrons, and as before $\hat{X} = \sum_{i=1}^N \hat{x}_i$. 
Resta\cite{resta98} showed that in the thermodynamic limit the 
electronic polarization is given by 

\begin{equation}
\label{P_resta}
P_{\rm el} = \lim_{N \rightarrow \infty}
\frac{q_e}{2 \pi} {\rm Im} \ln z_N.
\end{equation}

\noindent However, the
nature of the approximations involved at finite $L$, and the precise relation 
between Eqs. \ref{delta_P_om94} and \ref{P_resta} as a function
of the size of the system, were not clarified; this is a matter of crucial 
importance for the usefulness of the expressions in practical calculations,
and is discussed in Section \ref{comparison}.

Resta and Sorella\cite{resta99} proposed to measure the localization length in
1D insulators as\cite{silvestrelli}

\begin{equation}
\label{resta_lambda}
\xi = \sqrt{\cal D}/(2 \pi n_0), 
\end{equation}

\noindent where $n_0 = N/L$ and 

\begin{equation}
\label{loc_resta}
{\cal D} = - \lim_{N \rightarrow \infty} N \ln {\big| z_N \big|}^2.
\end{equation}

\noindent They showed that if the electrons are uncorrelated, then
for insulators
$\xi$ is simply related to the 1D version of $\Omega_{\rm I}$ 
defined in Eq. \ref{omega_I} ($\xi^2 = \Omega_{\rm I}/M$), and therefore
$\xi$ is finite, whereas for metals it diverges (even before taking the
limit $N \rightarrow \infty$). Then they proposed that a similar behavior 
should occur when the electrons are correlated (with the difference that in 
general $\xi$ diverges for correlated conductors  only after taking the limit).

In the present work we generalize Eq. \ref{resta_lambda} to many dimensions
(see Section \ref{discrete_fix}), and give an explicit many-body derivation 
that, similarly to $\Omega_{\rm I}$ in the uncorrelated case,
$\xi^2$ measures the polarization fluctuations in correlated systems. 
Similarly to the continuum formulas involving an average over twisted boundary
conditions\cite{aligia99a}, these formulas require modification when there is a
non-integer number of electrons per cell\cite{aligia99b} (see 
Appendix \ref{frac_fill}).

\section{GENERATING FUNCTION FORMALISM}\label{moment_gen_fnct}

\subsection{Definitions}\label{definitions}

Generating functions play a central role in the theory of 
statistics\cite{kendall}, and have been applied to many problems in 
physics\cite{kubo62,fulde}. Loosely speaking, a generating
function of a distribution is some function which yields, upon successive 
differentiation, the moments of the distribution, or some combination thereof.
Two kinds of generating functions will be of interest to us: the 
{\it characteristic function} $C_{\bf X}({\boldsymbol \alpha})$, and its 
logarithm, the {\it cumulant generating function}. If ${\bf X}$ is a vector of 
$d$ variables $X_1,...,X_d$ with a normalized joint probability distribution 
function $p(X_1,...,X_d)=p({\bf X})$, the characteristic function is defined as

\begin{equation}
\label{characteristic_function}
C_{\bf X}({\boldsymbol \alpha}) = \int_{- \infty}^{+ \infty}
e^{-\imath {\boldsymbol \alpha} \cdot {\bf X}} p(X_1,...,X_d) dX_1...dX_d
\equiv \left< e^{-\imath {\boldsymbol \alpha} \cdot {\bf X}} \right>,
\end{equation} 

\noindent where ${\boldsymbol \alpha} \cdot {\bf X} = 
\sum_{i=1}^d \alpha_i X_i$\cite{foot_average}. The $d$-dimensional moments can
be extracted directly from $C_{\bf X}({\boldsymbol \alpha})$:

\begin{equation}
\label{moments}
{\left< X_1^{n_1} ... X_d^{n_d} \right>} = 
\imath^n {\left. \partial^n_{\alpha_1^{n_1} ... \alpha_d^{n_d}}
C_{\bf X}({\boldsymbol \alpha}) 
\right|}_{{\boldsymbol \alpha}={\bf 0}},
\end{equation} 

\noindent where $n = \sum_{i=1}^d n_i$.
The cumulants are obtained in a similar way from 
$\ln C_{\bf X}({\boldsymbol \alpha})$:

\begin{equation}
\label{cumulants}
{\left< X_1^{n_1} ... X_d^{n_d} \right>}_c = 
\imath^n {\left. \partial^n_{\alpha_1^{n_1} ... \alpha_d^{n_d}}
\ln C_{\bf X}({\boldsymbol \alpha}) \right|}_{{\boldsymbol \alpha}={\bf 0}},
\end{equation} 

\noindent where, following the notation of Ref. \cite{kubo62}, 
${\left< ... \right>}_c$ denotes the {\it cumulant average}, which in
general is different from the simple average $\left< ... \right>$ associated 
with the moments: 

\begin{eqnarray}
\label{first_cumulants}
&{\left< X_i \right>}_c = \left< X_i \right> \nonumber \\
&{\left< X_i^2 \right>}_c = \left< X_i^2 \right> - {\left< X_i \right>}^2
\nonumber \\
&{\left< X_i X_j \right>}_c = \left< X_i X_j \right> - 
\left< X_i \right> \left< X_j \right>.
\end{eqnarray}

An important property of cumulants is that they can be explicitly
represented solely in terms of the lower moments, and vice-versa. 
More precisely - and this is very relevant for what follows - for $n>1$
they can be expressed in terms of the {\it central moments} 
$\left< {\Delta X}_1^{m_1} {\Delta X}_2^{m_2} {\Delta X}_3^{m_3} \right>$, 
where ${\Delta X}_i = X_i - \left< X_i \right>$ and $m_1+m_2+m_3 \leq n$,
and thus they are independent of the mean $\left< X_i \right>$.
Moreover, provided that the characteristic function exists, the set of all the 
moments or cumulants completely determines the distribution. 

\subsection{Polarization distribution in finite systems}
\label{mgf_dipole}

Let us consider a neutral 3D system of finite volume $V$ containing $N$
electrons and $N_{\rm n}$ nuclei. The dipole moment operator is 
$\hat{\bf d} = q_{\rm e} \hat{\bf X} + q_{\rm n} \hat{\bf X}_{\rm n}$, where
$\hat{\bf X} = \sum_{i=1}^N \hat{\bf x}_i$ and
$\hat{\bf X}_{\rm n} = \sum_{i=1}^{N_{\rm n}} \hat{\bf x}^{\rm n}_i$.
The average dipole moment of the system is
${\bf d} = \left< \Psi \left| \hat{\bf d} \right| \Psi \right>=
\int {\bf r} \rho({\bf r}) d {\bf r}$. Since the center of mass ${\bf X}/N$ of
the electrons is not perfectly localized, $\Psi$ is not an eigenstate of
$\hat{\bf X}$. Therefore the cartesian components of the dipole 
moment undergo quantum fluctuations, having a joint probability distribution 
$p({\bf d})$ dictated by the ground state wavefunction.
For simplicity we will assume that the nuclei can be treated classically as 
``clamped'' point charges; then they only contribute to the average of the 
distribution, and the quantum fluctuations come solely from the electrons. 
Hence, in what follows we will neglect the nuclear contribution, focusing on the 
distribution of the electronic center of mass. If $\Psi$ is the many-electron
wavefunction (parametrized by the nuclear coordinates) with normalization 
$\left< \Psi \big| \Psi \right> = 1$, that distribution is given by

\begin{equation}
\label{X_dist}
p({\bf X}) = 
\left< \Psi \left| \delta(\hat{\bf X} - {\bf X}) \right| \Psi \right>.
\end{equation}

\noindent Similarly, for a given component, say, $X_1$, the distribution is

\begin{equation}
\label{X_1_dist}
p(X_1) = \int_{- \infty}^{+ \infty} p(X_1,X_2,X_3) d X_2 d X_3 =
\left< \Psi \left| \delta(\hat{X}_1 - X_1) \right| \Psi \right>.
\end{equation}

\noindent The characteristic function $C_{\bf X}({\boldsymbol \alpha})$ is 
obtained by substituting Eq. \ref{X_dist} into 
Eq. \ref{characteristic_function}:

\begin{equation}
\label{char_fnct_finite}
C_{\bf X}({\boldsymbol \alpha}) = \left< \Psi \left| 
e^{-\imath {\boldsymbol \alpha} \cdot \hat{\bf X}} \right| \Psi \right>.
\end{equation}

It is clear that if we define
$\left< {\hat{X}_1}^{n_1} {\hat{X}_2}^{n_2} {\hat{X}_3}^{n_3} \right> = 
\left< \Psi \left| {\hat{X}_1}^{n_1} {\hat{X}_2}^{n_2} 
{\hat{X}_3}^{n_3} \right| \Psi \right>$ we find, using Eq. \ref{moments}, 
$\left< X_1^{n_1}  X_2^{n_2} X_3^{n_3} \right> =
\left< {\hat{X}_1}^{n_1} {\hat{X}_2}^{n_2} {\hat{X}_3}^{n_3} 
\right>$\cite{foot_average}.
The electronic polarization operator is
${\left( \hat{P}_{\rm el} \right)}_i = q_e \hat{X_i} / V$, and the moments of 
its distribution are given by

\begin{equation}
\label{moments_polarization}
\left< {\left( P_{\rm el}\right)}_1^{n_1}  {\left( P_{\rm el}\right)}_2^{n_2}
{\left( P_{\rm el}\right)}_3^{n_3} \right> = 
{\left( \frac{q_e}{V} \right)}^n  
\left< \hat{X_1}^{n_1} \hat{X_2}^{n_2} \hat{X_3}^{n_3}
\right>,
\end{equation}

\noindent where $n=n_1+n_2+n_3$.

\subsection{Polarization distribution in extended systems
}\label{gen_extended}

In the case of a finite system the characteristic function 
$C_{\bf X}({\boldsymbol \alpha})$ was introduced as a
purely formal device for obtaining the moments of the distribution, since in 
practice it was completely equivalent to a direct evaluation of the moments. In
the case of an extended system the situation is rather different since, as 
discussed in the Introduction, the very definition of polarization
needs to be reexamined, and a naive generalization of the direct method of 
calculating the moments does not apply.

Again let us consider a system with $N$ electrons in a volume $V$.
As in previous work described in Section \ref{recent_develop}, it will be 
convenient to use twisted boundary conditions (Eq. \ref{twisted}). The 
difficulties in defining the macroscopic bulk polarization can be seen from
the fact that $\hat{X_i}^n$ is not a valid operator in the Hilbert space 
defined by Eq. \ref{twisted}: indeed, if $| \Psi_{\bf k} >$ is a
vector in that space, $\hat{X_i}^n | \Psi_{\bf k} >$  is not (it is not even 
normalizable), and thus ${\left< \hat{X_i}^n \right>}_{\bf k} =  
\left< \Psi_{\bf k} \left| {\hat{X_i}}^n \right| \Psi_{\bf k} \right>$
is ill-defined.
We will now show that although the {\it operator}  $\hat{\bf X}$ is
ill-defined, one can nevertheless define a meaningful joint probability 
{\it distribution} $p({\bf X})$ for the electronic center of
mass in an extended insulating system. This will have the same physical 
interpretation as
Eq. \ref{X_dist} for a finite system, i.e., the moments of the variables 
${\left( P_{\rm el} \right)}_i = q_e X_i / V$ are the moments of the
distribution of the electronic polarization:

\begin{equation}
\label{moments_polarization_bulk}
\left< {\left( P_{\rm el}\right)}_1^{n_1}  {\left( P_{\rm el}\right)}_2^{n_2}
{\left( P_{\rm el}\right)}_3^{n_3} \right> = 
{\left( \frac{q_e}{V} \right)}^n  
\left< X_1^{n_1} X_2^{n_2} X_3^{n_3}
\right>.
\end{equation}

\noindent Our main interest will be in the average $\left< X_i \right>$ and in
the quadratic spread $\left< {\Delta X_i}^2 \right>$ which, according to
Ref. \cite{kudinov}, measures the electronic localization. More precisely, we
will define the localization length $\xi_i$ along the $i$-th direction as

\begin{equation}
\label{loc_length}
\xi_i^2= \lim_{N \rightarrow \infty} \frac{1}{N} \left< \Delta X_i^2 \right>.
\end{equation}

The desired distribution $p({\bf X})$ can be obtained via a proper 
generalization of the characteristic function 
$C_{\bf X}({\boldsymbol \alpha})$, which for a finite system was given by 
Eq. \ref{char_fnct_finite}. We start by introducing the quantity 
${\cal C}({\bf k},{\boldsymbol \alpha})$, which is a {\it precursor} to the 
characteristic function for the extended system:

\begin{equation}
\label{precursor_mgf}
{\cal C}({\bf k},{\boldsymbol \alpha}) = 
\left<\Psi_{\bf k} \left| e^{-\imath {\boldsymbol \alpha} \cdot  \hat{\bf X}} 
\right| \Psi_{{\bf k}+{\boldsymbol \alpha}} \right> =
\left< \left. \Phi_{\bf k} \right| \Phi_{{\bf k}+{\boldsymbol \alpha}}
\right>,
\end{equation}

\noindent where in the last equality we used Eq. \ref{phi}. In the first form 
the similarity to Eq. \ref{char_fnct_finite} is evident\cite{foot_dotproduct};
notice, however, that this is not an expectation value 
over a single ground state $\Psi_{\bf k}$, since the boundary conditions on the
{\it bra} (${\bf k}$) and on the {\it ket} (${\bf k} + {\boldsymbol \alpha}$) 
are different. That is required in order to compensate for the shift by 
$-{\boldsymbol \alpha}$ in the boundary condition on the {\it ket} caused by 
the operator $e^{-\imath {\boldsymbol \alpha} \cdot  \hat{\bf X}}$; in this way
the states $\left. \left| \Psi_{\bf k} \right. \right>$ and 
$\left. \left| \tilde{\Psi}_{\bf k} 
({\boldsymbol \alpha}) \right. \right> = 
e^{-\imath {\boldsymbol \alpha} \cdot  
\hat{\bf X}} \left. \left| \Psi_{{\bf k}+{\boldsymbol \alpha}} \right. \right>$
obey the same
boundary conditions even for $\alpha_i \not= 2 \pi n/L_i$, so that their 
dotproduct $\left< \Psi_{\bf k} \left| \tilde{\Psi}_{\bf k} 
({\boldsymbol \alpha}) \right. \right>$ does not vanish and
${\cal C}({\bf k},{\boldsymbol \alpha})$ can be chosen to be a differentiable
function (we are assuming that the ground state insulating wavefunction
$\Psi_{\bf k}$ is non-degenerate and is separated in energy from the excited 
states by a finite gap).

That ${\cal C}({\bf k},{\boldsymbol \alpha})$ is not yet the characteristic
function for the extended system can be seen from the fact that the first 
``moment'' that it generates is not gauge-invariant, as 
required from any physical quantity\cite{foot_quantum}. 
If ${\cal C}({\bf k},{\boldsymbol \alpha})$ were the characteristic function, 
the first moment would be, according to Eq. 
\ref{moments}, ${\left< X_i \right>}_{\bf k}=\imath {\left. \partial_{\alpha_i}
{\cal C}({\bf k},{\boldsymbol \alpha}) \right|}_{{\boldsymbol \alpha}=0} =
\imath \left< \Phi_{\bf k} \left| \partial_{k_i} \Phi_{\bf k} \right. \right>$.
But this is nothing other than the {\it Berry connection}, which is
a gauge-dependent quantity (see, for instance, Ref. \cite{resta_lect}).

\subsection{Gauge-invariant cumulants}
\label{gi_cumulant}

Gauge-invariant moments and cumulants for the extended system can be obtained 
from the following cumulant generating function, which is the central quantity
in this work: 

\begin{equation}
\label{bulk_M}
\ln C({\boldsymbol \alpha}) =
\frac{V}{{(2 \pi)}^3} \int d {\bf k}
\ln {\cal C}({\bf k},{\boldsymbol \alpha}),
\end{equation}

\noindent where the average is over all twisted boundary conditions (for 
special cases where the number of electrons per primitive cell is not an 
integer, see Appendix \ref{frac_fill}). The cumulants are then obtained in the
same way as in Eq. \ref{cumulants}:

\begin{equation}
\label{cumulants_ext}
{\left< X_1^{n_1} ... X_d^{n_d} \right>}_c = 
\imath^n {\left. \partial^n_{\alpha_1^{n_1} ... \alpha_d^{n_d}}
\ln C({\boldsymbol \alpha}) \right|}_{{\boldsymbol \alpha}={\bf 0}},
\end{equation} 

\noindent and likewise for the moments.
Before we continue, we point out that $C({\boldsymbol \alpha})$
is in general different from the function 

\begin{equation}
\label{m_w}
C_{\rm W}({\boldsymbol \alpha}) =
\frac{V}{{(2 \pi)}^3} \int d {\bf k}
{\cal C}({\bf k},{\boldsymbol \alpha}).
\end{equation}

\noindent This is also a characteristic function, but of a different, 
gauge-dependent distribution, whose interpretation and relation to 
$C({\boldsymbol \alpha})$ will be discussed in Section \ref{loc_config_space}.

As mentioned in Section \ref{recent_develop}, Eq. \ref{delta_P_om94} is most
easily interpreted when using a 
periodic gauge:
$\Psi_{{\bf k}+{\bf G}} = \Psi_{\bf k}$. Since we will recover that equation
starting from $\ln C({\boldsymbol \alpha})$, a periodic gauge will be assumed 
in what follows; thus what is meant here by gauge-invariance is 
invariance under transformations which preserve the 
phase in Eq. \ref{global_topology}. The most general $\varphi({\bf k})$ in 
Eq. \ref{gauge_trans} which complies with this requirement 
is\cite{ksv,weinreich}:

\begin{equation}
\label{varphi} 
\varphi({\bf k}) = \beta({\bf k}) - {\bf k} \cdot {\bf R},
\end{equation}

\noindent where ${\bf R}$ is an arbitrary lattice vector and 
$\beta({\bf k}+{\bf G})=\beta({\bf k})$.
The gauge-invariance of the cumulants generated by 
$\ln C({\boldsymbol \alpha})$ can now be seen as follows: according to Eqs. 
\ref{gauge_trans}, \ref{cumulants}, and \ref{bulk_M}, the cumulants change 
under a general gauge transformation as

\begin{equation}
\label{cumulant_gauge}
{\left< X_1^{n_1} X_2^{n_2} X_3^{n_3} \right>}_c \rightarrow
{\left< X_1^{n_1} X_2^{n_2} X_3^{n_3} \right>}_c + \imath^{n+1}
\frac{V}{{(2 \pi)}^3} \int d {\bf k} 
\partial^n_{k_1^{n_1} k_2^{n_2} k_3^{n_3}} \varphi({\bf k}).
\end{equation}

\noindent 
Substitution of Eq. \ref{varphi} into
Eq. \ref{cumulant_gauge} shows that for $n=1$ the cumulants change as
$\left< {\bf X} \right> \rightarrow \left< {\bf X} \right> + {\bf R}$, whereas
for $n>1$ they remain unchanged. The change in $\left< {\bf X} \right>$ but not
in the higher cumulants (which do not depend on the mean) indicates a rigid 
shift by ${\bf R}$ of the whole distribution
$p({\bf X})$, and is related to the quantum of polarization\cite{ksv,om94} 
(see Appendix \ref{quantum}).

The cumulants are also real, as expected:  defining
$f_{n_1n_2n_3}({\bf k}) = \imath^n {\left.
\partial^n_{\alpha_1^{n_1} \alpha_2^{n_2} \alpha_3^{n_3}} 
\ln \left< \Phi_{\bf k} \left| \Phi_{{\bf k}+{\boldsymbol \alpha}} \right.
\right> \right|}_{{\boldsymbol \alpha}=0}$ and
${\boldsymbol \beta} = - {\boldsymbol \alpha}$, we obtain
$f_{n_1n_2n_3}(-{\bf k})= {(-1)}^n \imath^n 
\partial^n_{\beta_1^{n_1} \beta_2^{n_2} \beta_3^{n_3}}
\ln \left< \Phi_{-{\bf k}} \big| \Phi_{-({\bf k}+{\boldsymbol \beta})} 
\right>\big|_{{\boldsymbol \beta}=0}$. Choosing 
a gauge such that $\Phi_{- {\bf k}} = \Phi_{\bf k}^*$, which can always be done
due to time-reversal symmetry\cite{weinreich}, leads to
$\left< \Phi_{-{\bf k}} \left| \Phi_{-({\bf k}+{\boldsymbol \beta})} \right.
\right> = 
\left< \left. \Phi_{{\bf k}+{\boldsymbol \beta}} \right| \Phi_{\bf k} \right>$,
and thus $f_{n_1n_2n_3}(-{\bf k}) = {\left[ f_{n_1n_2n_3}({\bf k}) \right]}^*$.
That completes the
proof, since according to Eqs. \ref{precursor_mgf}, \ref{bulk_M}, and 
\ref{cumulants_ext}, ${\left< X_1^{n_1} X_2^{n_2} X_3^{n_3} \right>}_c$ is 
given by the average over ${\bf k}$ of $f_{n_1n_2n_3}({\bf k})$.
(Since gauge-invariance was previously established, the choice
$\Phi_{- {\bf k}} = \Phi_{\bf k}^*$ can be made without loss of generality.)

Let us now look at explicit expressions for the first few cumulants.
Combining Eqs. \ref{precursor_mgf} and \ref{bulk_M} and taking the first
derivative, we find

\begin{equation}
\label{first_cumulant_derivative} {\left.
\partial_{\alpha_i} \ln C({\boldsymbol \alpha}) 
\right|}_{{\boldsymbol \alpha}=0} = 
\frac{V}{{(2 \pi)}^3} \int d{\bf k}
\left< \Phi_{\bf k} \left| \partial_{k_i} \Phi_{\bf k} \right. \right>.
\end{equation}

\noindent Together with Eq. \ref{cumulants}, this gives

\begin{equation}
\label{first_cumulant}
{\left< X_i \right>}_c = \frac{\imath V}{{(2 \pi)}^3} \int d{\bf k}
\left< \Phi_{\bf k} \left| \partial_{k_i} \Phi_{\bf k} \right. \right>,
\end{equation}

\noindent which, with the help of Eq. \ref{moments_polarization_bulk}, is seen
to be precisely the Berry phase expression for the average electronic
polarization (Eq. \ref{delta_P_om94}). Our formulation in terms of
$C({\boldsymbol \alpha})$ therefore agrees with the Berry phase theory 
of polarization\cite{ksv,vks,om94}. It is however more general, since it also 
provides the higher moments: for $n=2$, similar steps as before lead to

\begin{equation}
\label{second_cumulant}
{\left< X_i X_j \right>}_c = - \frac{V}{{(2 \pi)}^3} \int d{\bf k}
\left[
\left< \Phi_{\bf k} \left| \partial^2_{k_i k_j} \Phi_{\bf k} \right. \right> -
\left< \Phi_{\bf k} \left| \partial_{k_i} \Phi_{\bf k} \right. \right>
\left< \Phi_{\bf k} \left| \partial_{k_j} \Phi_{\bf k} \right. \right>
\right].
\end{equation}

\noindent Integrating the first term by parts (using the periodic 
gauge condition) and noting that
$\left< \Phi_{\bf k} \left| \partial_{k_i} \Phi_{\bf k} \right. \right> = 
- \left< \left. \partial_{k_i} \Phi_{\bf k} \right| \Phi_{\bf k} \right>$, this
becomes
${\left< X_i X_j \right>}_c = V/(8 \pi^3) \int d{\bf k}
T_{ij}({\bf k})$, where

\begin{equation}
\label{geometric_tensor}
T_{ij}({\bf k}) = \left< \left. \partial_{k_i} \Phi_{\bf k} \right|
\partial_{k_j} \Phi_{\bf k} \right> - 
\left< \left. \partial_{k_i} \Phi_{\bf k} \right| \Phi_{\bf k} \right>
\left< \Phi_{\bf k} \left| \partial_{k_j} \Phi_{\bf k} \right. \right>
\end{equation}

\noindent is the gauge-invariant 
{\it quantum geometric tensor}\cite{berry89} [see also Eq. C9
of Ref. \cite{marzari97}]. The real part of $T_{ij}({\bf k})$ is the metric 
tensor ${G}_{ij}({\bf k})$ first introduced by Provost and 
Vallee\cite{provost}:

\begin{equation}
\label{big_metric}
{G}_{ij}({\bf k}) = 
{\rm Re}  \left< {\partial}_{k_i} \Phi_{\bf k} \left|
{\partial}_{k_j} \Phi_{\bf k} \right. \right> -
\left< \left. {\partial}_{k_i} \Phi_{\bf k} \right| 
 \Phi_{\bf k} \right>
  \left< \Phi_{\bf k} \left| {\partial}_{k_j} 
\Phi_{\bf k} \right. \right>,
\end{equation}

\noindent where the second term is automatically real. ${G}_{ij}({\bf k})$
is the many-body analogue of the tensor $g_{ij}({\bf k})$ defined in
Eq. \ref{metric}. Using Eq. \ref{first_cumulants}  and the fact that the cumulants
are real, ${\left< X_i X_j \right>}_c$ can be rewritten as

\begin{equation}
\label{second_cumulant_c}
{\left< X_i X_j \right>} - \left< X_i \right> \left< X_j \right> = 
\frac{V}{{(2 \pi)}^3} \int d{\bf k}
G_{ij}({\bf k}),
\end{equation}

\noindent which becomes, after taking the trace, the many-body counterpart of
Eq. \ref{omega_I}. The above equation establishes the physical interpretation 
of the ${\bf k}$-averaged metric tensor as the mean-square fluctuation 
of the macroscopic bulk polarization (see also the next 
Section). The connection between such ``quantum metrics'' and quantum 
fluctuations was pointed out already in Ref.\cite{provost}.

\section{fluctuation-dissipation relation}
\label{fdt}

We have seen that the first cumulant agrees with the Berry phase expression for
the average macroscopic 
polarization. Here we will show that the expression for the second cumulant,
Eq. \ref{second_cumulant_c}, is consistent with the fluctuation-dissipation
relation\cite{callen,callen52,landau} between the bulk polarization 
fluctuations and the optical conductivity
${\rm Im} {\chi}_{ij}(\omega)=(1/\omega) {\rm Re} \sigma_{ij}(\omega)$.

For a {\it finite} system with open boundary conditions, which is the case
discussed by Kudinov\cite{kudinov}, the Kubo-Greenwood formula for the 
conductivity can be written in terms of the off-diagonal position matrix 
elements, $X_{nm}^i = \left< \Psi_n \left| \hat{X}_i \right| \Psi_m \right>$:

\begin{equation} 
\label{kubo_greenwood_finite}
{\rm Re} \sigma^n_{ij}(\omega) =
\frac{\pi q_e^2}{m_e^2 \hbar \omega V} \sum_{m \not= n}
\omega^2_{mn} {\rm Re}
\Bigl[
X_{nm}^i X_{mn}^j \delta \left( \omega_{mn} - \omega  \right) - 
X_{nm}^j X_{mn}^i \delta \left( \omega_{mn} + \omega \right)
\Bigr],
\end{equation}

\noindent where $\hbar \omega_{mn} = E_m - E_n$. Alternatively, the well-known
relation 

\begin{equation}
\label{commutation_xp_finite}
P^i_{nm} = \imath m_e  \omega_{nm} X^i_{nm}
\hspace{0.1in} (m \not= n)
\end{equation}

\noindent can be used to rewrite Eq. \ref{kubo_greenwood_finite} in terms of 
momentum matrix elements. Whereas for a finite system the two formulas are 
interchangeable, for an extended system the position matrix elements become 
ill-defined, and therefore only the latter form remains valid.

At this point it is convenient to introduce the notation

\begin{equation}
\label{pos_matrix}
X_{nm,\bf k}^i = \imath \left< \Phi_{n \bf k} \left| {\partial}_{k_i} 
\Phi_{m \bf k} \right. \right> = {\left( X_{mn,\bf k}^i \right)}^*.
\end{equation}

\noindent The single-body analogue of such quantities is discussed in
Refs. \cite{blount,zak}. Our motivation for introducing them is the
following: if 
$\hat{\bf P}= \sum_{i=1}^N {\bf p}(i)$ is the many-body momentum operator for
the extended system, and
${\bf P}_{nm,\bf k} = 
\left< \Psi_{n{\bf k}} \left| \hat{\bf P} \right| \Psi_{m{\bf k}} \right>$,
in the case of a non-relativistic hamiltonian without spin-orbit coupling
Eqs. \ref{phi} and \ref{h_k} lead to

\begin{equation}
\label{commutation}
P_{nm,\bf k}^i =
\frac{m_e}{\hbar} \left< 
\Phi_{n \bf k} \left|
\left[ {{\partial}_k}_i, \hat{H}({\bf k}) \right] \right| 
\Phi_{m \bf k} \right>
= \imath m_e  \omega_{nm}({\bf k}) X^i_{nm,\bf k}
\hspace{0.1in} (m \not= n),
\end{equation}

\noindent which is formally identical to Eq. \ref{commutation_xp_finite}.
Notice, however, that we are now dealing
with an {\it extended} system with twisted boundary conditions, for which the
proper position matrix elements, 
$\left< \Psi_{n \bf k} \left| \hat{X}_i \right| \Psi_{m \bf k} \right>$, are 
ill-defined; the above relation shows that they should be replaced
by the quantities $X_{nm,\bf k}^i$. Substituting Eq. \ref{commutation} into the
Kubo-Greenwood formula in the form valid for an extended system, we are left 
with an expression for ${\rm Re} \sigma^{n \bf k}_{ij}(\omega)$ formally 
identical to Eq. \ref{kubo_greenwood_finite}, with $X_{nm}^i$ replaced by
$X_{nm,\bf k}^i$ and $\omega_{mn}$ replaced by $\omega_{mn}({\bf k})$.

Let us now specialize to the ground state ($n=0$), assuming that it is
non-degenerate; then $\omega_{m0}({\bf k})>0$, 
and we obtain

\begin{eqnarray}
\label{frequency_integral}
& \int_0^{\infty} \frac{d \omega}{\omega} 
{\rm Re} \sigma^{0 \bf k}_{ij}(\omega) =
\frac{\pi q_e^2}{\hbar V} {\rm Re} \left\{ \sum_{m > 0}
X^i_{0m,{\bf k}} X^j_{m0,{\bf k}} \right\} \nonumber \\
& = \frac{\pi q_e^2}{\hbar V} {\rm Re} \left\{
\left< {{\partial}_k}_i \Phi_{0 \bf k} \left| 
\left( \sum_{m > 0} 
\right| \Phi_{m \bf k} \right>
\left< \Phi_{m \bf k} \left| \right)
\right| {{\partial}_k}_j  \Phi_{0 \bf k} \right> \right\}.
\end{eqnarray}

\noindent Using the completeness relation
to eliminate the excited states on the r.h.s, and comparing with 
Eq. \ref{geometric_tensor}, we find
$T_{ij}({\bf k}) = \sum_{m>0} X^i_{0m,{\bf k}} X^j_{m0,{\bf k}}$.
Together with $G_{ij}({\bf k})={\rm Re} T_{ij}({\bf k})$, this yields

\begin{equation}
\label{fdt_bulk_single_k}
\int_0^{\infty} \frac{d \omega}{\omega} 
{\rm Re} \sigma^{0 \bf k}_{ij}(\omega) =
\frac{\pi q_e^2}{\hbar V} G_{ij}({\bf k}).
\end{equation}

\noindent Averaging over all twisted boundary
conditions and using Eqs. \ref{cumulants_ext} and \ref{second_cumulant_c}, we
arrive at the desired relation:

\begin{equation}
\label{fdt_bulk}
-\frac{1}{N} {\left. \partial^2_{\alpha_i \alpha_j} 
\ln C({\boldsymbol \alpha}) \right|}_{{\boldsymbol \alpha}=0} =
\frac{1}{N} \left[
{\left< X_i X_j \right>} - \left< X_i \right> \left< X_j \right> \right] = 
\frac{\hbar}{\pi q_e^2 n_0} \int_0^{\infty} \frac{d \omega}{\omega}
{\rm Re} \overline{\sigma}_{ij}(\omega),
\end{equation}

\noindent where $n_0 = N/V$ and  
$\overline{\sigma}_{ij}(\omega) = (V/8\pi^3)
\int d{\bf k} \sigma^{0 \bf k}_{ij}(\omega)$. 
Eq. \ref{fdt_bulk} is precisely the fluctuation-dissipation relation for more 
than one variable\cite{callen52,landau} at $T=0$ (compare with Eq. \ref{fdt1}).

\section{Localization length}
\label{fluct_ins_cond}

\subsection{Relation to the conductivity}
\label{fluct_ins_cond_a}

Here we will generalize to extended systems
Kudinov's analysis of the dipole moment fluctuations as a way to distinghish
insulators from conductors\cite{kudinov}, and discuss it in terms of a
localization length. It is convenient for that purpose to classify solids into 
three categories, according to the low-frequency behavior of the 
conductivity at $T=0$ as $V \rightarrow \infty$:

\begin{equation}
\label{real_cond_low_freq}
\left\{ \begin{array}{ll}
\lim_{\omega \rightarrow 0} {\rm Re} \sigma(\omega) = 0
&\mbox{insulators}\\
{\rm Re} \sigma(\omega) = (2 \pi q_e^2/\hbar^2) D \delta(\omega) + {\rm Re} 
\sigma_{\rm reg}(\omega)&\mbox{ideal conductors 
} \\ 
\lim_{\omega \rightarrow 0} {\rm Re} \sigma(\omega) = \sigma_0   
&\mbox{non-ideal conductors 
} \\
\end{array} 
\right.
\end{equation}

\noindent Insulators are characterized by a vanishing dc conductivity, in 
contrast to conductors. The singular contribution 
$(2 \pi q_e^2/\hbar^2)D\delta(\omega)$ occurs in ideal conductors - those 
without any scattering mechanism - and $D$ is called the ``Drude weight'' or 
``charge stiffness''\cite{fulde,scalapino}.
If there is scattering, the $\delta$ function peak is smeared out to a 
Lorentzian, so that $D=0$ and the dc conductivity of non-ideal conductors, such
as disordered metallic alloys, has a finite value $\sigma_0$;
$\sigma_{\rm reg}(\omega)$ is the regular finite-frequency part of 
$\sigma(\omega)$ in perfect conductors.

The only possible divergence of the integral on the r.h.s. of Eq. \ref{fdt_bulk} 
is around $\omega = 0$\cite{foot_divergence}, because as $\omega \rightarrow \infty$
${\rm Im} \chi_{ii}(\omega) \sim \omega^{-3}$\cite{jackson}.
Substituting Eq. \ref{real_cond_low_freq} into Eq. \ref{fdt_bulk} we see that 
the quantity $\xi_i$ in Eq. \ref{loc_length}
is finite for insulators and diverges for conductors.
By the same token, using the zero-frequency limit of the Kramers-Kr\"onig 
relation\cite{landau_em,jackson}, one finds the familiar result that the static
susceptibility is finite for insulators and divergent for conductors.
Hence, assuming Eq. \ref{real_cond_low_freq}, at $T = 0$
the following three conditions are equivalent: 
(i) ${\rm Re} \chi_{ii}(0)$ is finite, (ii) 
$\lim_{\omega \rightarrow 0} {\rm Re} \sigma_{ii}(\omega) = 0$, and
(iii) $\xi_i$
is finite. 

The quantity $\xi_i$ has the dimensions of length. Since it is finite for 
insulators and infinite for conductors, it is natural to interpret it
as an electronic localization length along the $i$-th direction. 
According to Eq. \ref{fdt_bulk}, for extended systems it can be written as 
$\xi_i = \lim_{N \rightarrow \infty} \xi_i(N)$, where $\xi_i(N)$ is  given in 
Table I in terms of the cumulant generating function.
This formula has a striking similarity to the Drude weight formula derived by 
Kohn in terms of the total energy $E({\bf k})$\cite{kohn,scalapino,stafford91},
given also in Table I:
both are second derivatives of some quantity with respect to the twisted 
boundary conditions (in the case of $\xi_i(N)$ the twisting of the boundary 
condition ${\bf k}$ is followed by an averaging over all ${\bf k}$, 
hence the parameter ${\boldsymbol \alpha}$ instead of ${\bf k}$). 

It is clear from Table I that unlike the localization length, the Drude weight
does not provide a universal criterion to discern insulators from 
conductors\cite{kohn,kohn68}. However, the combination of the two 
quantities in principle enables us to distinguish between the three categories.
In the same way that the Drude weight measures the ``degree of conductivity''
of an ideal conductor, 
$1 / \xi_i$ measures the degree of localization of the
electrons in an insulator. Insofar as localization - in a properly defined 
sense - is an essential property of the insulating state\cite{kohn,kohn68}, 
this can be viewed as a meaningful measure of the ``degree of insulation'', one
which is expected to apply to all types of insulators.
 
Although we have managed to express 
$\xi_i$ in terms of the measurable optical conductivity via the 
fluctuation-dissipation relation, it is not yet clear how it relates to the 
notion of localization put forward by Kohn, in terms of the localization 
properties of the insulating wavefunction in configuration space. That will be
discussed in Section \ref{loc_config_space}.

\subsection{Relation to the optical gap}

In an insulator optical absorption starts at a threshold energy $E_{\rm g}$,
below which ${\rm Re} \sigma_{ii}(\omega)=0$ (we are neglecting
phonon-assisted transitions, so that the gap $E_{\rm g}$ is the minimum
gap for optical transitions). From this it follows that

\begin{equation}
\label{inequality}
\int_0^{\infty} \frac{d \omega}{\omega}{\rm Re} \sigma_{ii}(\omega) \leq
\frac{\hbar}{E_{\rm g}} \int_0^{\infty} d \omega {\rm Re} \sigma_{ii}(\omega).
\end{equation} 

\noindent With the help of the sum rule for oscillator
strengths\cite{landau_em,jackson,bassani}
$\int_0^{\infty} d \omega {\rm Re} \sigma_{ii}(\omega) = 
(1/8) \omega_{\rm p}^2$ ($\omega_{\rm p}$ is the plasma frequency), together 
with Eqs. \ref{loc_length} and \ref{fdt_bulk}, we conclude that

\begin{equation}
\label{wf_spread_gap}
\xi_i^2 \leq \frac{\hbar^2}{2 m_e E_{\rm g}}.
\end{equation}

\noindent This inequality shows that the polarization fluctuations are 
controlled by the optical gap, lending support to the intuitive notion that the
larger the gap, the more localized the electrons are. It strongly resembles an
inequality previously derived by Kivelson\cite{kivelson} for non-interacting
electrons in 1D, where $\xi_i^2$ is replaced by the quadratic spread of 
properly chosen Wannier functions. As discussed in 
Appendix \ref{omega_I_fluct}, in fact in 1D $\xi_i^2$ equals the average spread
of the maximally-localized Wannier functions.

\section{``many-body wannier functions''}
\label{connection_kohn}

The expression for the cumulant generating function, Eq. \ref{bulk_M}, involves
an average over all twisted boundary conditions, which was introduced in a 
somewhat {\it ad hoc} manner in order to render the resulting distribution
gauge-invariant (modulo a rigid shift by a quantum). We will now shed some 
light on the physical significance of the averaging procedure, by showing how 
it can be rationalized in terms of the 
notion of localization in insulators developed by Kohn\cite{kohn,kohn68}. This
will be achieved by introducing ``many-body Wannier functions'', and will allow
us to tie together the Berry phase theory of polarization and Kohn's theory of
localization.

\subsection{Finite system: Localization in real space}
\label{loc_real_space}

It is instructive to start by discussing the 
case of a finite system
of linear dimensions $\sim a$ (e.g., a molecule
with $N$ electrons). Instead of the usual open boundary conditions, we can
choose to impose periodic boundary conditions
on its wavefunction  (${\bf k}=0$ in Eq. \ref{twisted}), 
choosing $L>>a$. The resulting $N$-electron wavefunction is periodic in 
configuration space, as depicted schematically in Fig. 1 for 1D and $N=2$; it
is a sum of partial wavefunctions which only differ from one another by a 
translation of the coordinates,

\begin{equation}
\label{pbc_molecule_b}
\Psi_{{\bf k}=0}({\bf x}_1,...,{\bf x}_N) = \sum_{{\bf m}_1} ... 
\sum_{{\bf m}_N} \Psi_{{\bf m}_1 ... {\bf m}_N}({\bf x}_1,...,{\bf x}_N),
\end{equation}

\noindent where the $\{{\bf m}_i\}$ label the partial wavefunctions. These are
localized in geometrically equivalent regions 
${\cal R}_{{\bf m}_1...{\bf m}_N}$ in configuration space (shaded regions in 
Fig. 1), which for $L>>a$ are essentially non-overlapping:

\begin{equation}
\label {loc_molecule}
\Psi_{{\bf m}_1 ... {\bf m}_N} \Psi_{{\bf m}^{'}_1 ... {\bf m}^{'}_N}
\doteq 0, \hspace{0.2in} \mbox{for  }
({\bf m}_1, ..., {\bf m}_N) \not= ({\bf m}^{'}_1, ..., {\bf m}^{'}_N),
\end{equation}

\noindent where, using the notation of Refs. \cite{kohn,kohn68}, the symbol 
$\doteq$ denotes equality apart from exponential small corrections which vanish
in a manner such as $e^{-L/\xi}$, where 
$\xi$ is a localization length (in this example $\xi \sim a$). 

Next we switch from periodic to twisted boundary conditions (${\bf k} \not= 0$
in Eq. \ref{twisted}). From the localization properties it follows [see 
Section 2 of Ref. \cite{kohn} and Section VI of Ref. \cite{kohn68}] that the 
periodic part of the wavefunction, 
$\Phi_{\bf k} = e^{-\imath {\bf k} \cdot {\bf X}} \Psi_{\bf k}$, can be written
with an exponentially small error as

\begin{equation}
\label{twisted_molecule}
\Phi_{\bf k}({\bf x}_1,...,{\bf x}_N) \doteq e^{- \imath {\bf k} \cdot {\bf Q}}
\Phi_{{\bf k}=0}({\bf x}_1,...,{\bf x}_N),
\end{equation} 

\noindent where we have introduced the quantity

\begin{equation}
\label{pos_clausius}
{\bf Q} = \left\{ \begin{array}{ll}
\sum_{i=1}^N ({\bf x}_i - {\bf L}_{{\bf m}_i}) \mbox{ in } 
{\cal R}_{{\bf m}_1 ... {\bf m}_N} \\
{\bf F}({\bf x}_1,...,{\bf x}_N) \mbox{ outside all } 
{\cal R}_{{\bf m}_1 ... {\bf m}_N} \\
\end{array},
\right.
\end{equation}

\noindent where ${\bf L}_{{\bf m}_i}=L {\bf m}_i$, and ${\bf F}$ is a largely 
arbitrary, periodic function, which
joins smoothly with the values of ${\bf Q}$ at the boundaries of the regions
${\cal R}_{{\bf m}_1 ... {\bf m}_N}$.
Let us now look at
the precursor characteristic function 
${\cal C}({\bf k},{\boldsymbol \alpha})$ for such a 
finite system obeying twisted boundary conditions:
substituting the previous expressions into Eq. \ref{precursor_mgf}, we find

\begin{equation}
\label{precursor_mgf_molecule}
{\cal C}({\bf k},{\boldsymbol \alpha}) \doteq 
\left< \Phi_{{\bf k}=0} \left| 
e^{- \imath {\boldsymbol \alpha} \cdot \hat{\bf Q}}
\right| \Phi_{{\bf k}=0} \right> \doteq
\left< \Psi_{0 ... 0} \left| 
e^{- \imath {\boldsymbol \alpha} \cdot \hat{\bf X}}
\right| \Psi_{0 ... 0} \right>,
\end{equation}

\noindent which has the form of Eq. \ref{char_fnct_finite}, obtained using
open boundary conditions. It is clear that
as a result of localization of the electrons in real space the choice of
boundary conditions is immaterial. In particular,
${\cal C}({\bf k},{\boldsymbol \alpha})$ becomes independent of ${\bf k}$,
and thus from Eqs. \ref{bulk_M} and \ref{m_w} we conclude that 

\begin{equation}
\label{clausius_mossotti}
{\cal C}({\bf k},{\boldsymbol \alpha}) \doteq C_{\rm W}({\boldsymbol \alpha}) 
\doteq C({\boldsymbol \alpha}).
\end{equation}

\noindent Hence, averaging over boundary conditions becomes redundant, and
${\cal C}({\bf k},{\boldsymbol \alpha})$ is already the correct, 
gauge-invariant characteristic function. 
Finally, it is interesting to note that 
Eqs. \ref{twisted_molecule} and \ref{precursor_mgf_molecule} can be viewed
as a particular realization of Eqs. 2.22 and 2.23 of Ref. \cite{provost},
which were used to relate the quantum metric to the quantum fluctuations.

\subsection{Insulators: Localization in configuration space}
\label{loc_config_space}

Let us now consider an extended system, such as a crystal with
twisted  boundary conditions over a large volume $V$.
Although in general the charge density will be delocalized in real space,
Kohn has argued that if the system is insulating
the wavefunction is localized in configuration space\cite{kohn,kohn68}. 
Kohn's notion 
of localization is weaker than the one implied by Eq. \ref{loc_molecule} and 
Fig. 1; however, as we will see, it is sufficient to recover 
Eq. \ref{clausius_mossotti}, after making a judicious choice of gauge.

We start by introducing a
localized description of the many-electron insulating wavefunction in 
configuration space. In the non-interacting case, a localized one-electron 
description (in real space) is provided by the Wannier 
functions\cite{blount,wannier37,weinreich}. The Wannier function 
$\left. \left| {\bf R} n \right. \right>$ 
associated with band $n$ and centered around the unit cell labelled by the 
lattice vector ${\bf R}$ is related to the Bloch functions by the following 
unitary transformation:

\begin{equation}
\label{wannier}
\left. \left| {\bf R}n \right. \right>=
\frac{1}{\sqrt{N_c}}\sum_{\bf k} e^{-\imath{\bf k}\cdot{\bf R}}
\left. \left| \psi_{n{\bf k}} \right. \right>,
\end{equation}

\noindent where a periodic gauge is assumed, and the sum is over a uniform
grid of $N_c$ points in the Brillouin zone\cite{foot_continuum}. 
Due to the discretization of the integral over ${\bf k}$, 
$\left. \left| {\bf R}n \right. \right>$ 
is actually periodic in a large cell of volume $N_c v$ ($v$ is the
volume of the unit cell). The many-body analogue of a periodic Wannier function
(``many-body Wannier function'') can be defined in a similar way in terms of 
the many-body wavefunction $\Psi_{\bf k}$ (again assuming a periodic gauge):

\begin{equation}
\label{wannier_many_body_1}
\left. \left| W_{\bf M} \right. \right>=
\frac{1}{\sqrt{N_c}} \sum_{\bf k} e^{-\imath{\bf k}\cdot
{\bf R}_{\bf M}} \left. \left| \Psi_{\bf k} \right. \right>, 
\end{equation}

\noindent so that, using Eq. \ref{phi},

\begin{equation}
\label{phi_wf}
\left. \left| \Phi_{\bf k} \right. \right> = \frac{1}{\sqrt{N_c}} 
\sum_{\bf M}e^{-\imath{\bf k}\cdot 
(\hat{\bf X} - {\bf R}_{\bf M})} \left. \left| W_{\bf M} \right. \right>.
\end{equation}

\noindent Here ${\bf R}_{\bf M} = \sum_{i=1}^3 M_i {\bf L}_i$, where the $M_i$
are integers, and the vectors $\{ {\bf L}_1,{\bf L}_2,{\bf L}_3 \}$ define the
volume $V$ containing $N$ electrons (notice that $V$ is the volume of a 
supercell, typically a large multiple of the unit cell volume $v$). 
$\Phi_{\bf k}$ is periodic and $\Psi_{\bf k}$ obeys ${\bf k}$ boundary 
conditions over the volume $V^N$ in configuration space, and both 
$\Psi_{\bf k}$ and $W_{\bf M}$ are periodic over the volume $(N_c V)^N$. The 
normalization conventions are  the following: $\Phi_{\bf k}$ is normalized to 
one over a volume $V^N$, and $W_{\bf M}$ is normalized to one over a volume 
$(N_cV)^N$. Similarly to the 
$\left. \left| {\bf R}n \right. \right>$,
the $\left. \left| W_{\bf M} \right. \right>$ form an
orthonormal set.

From the same type of general considerations which are used to show that 
Wannier functions can be chosen localized\cite{wannier37}, it 
follows that we can choose $W_{\bf M}$ localized in the variable 
${\bf X}=\sum_{i=1}^N{\bf x}_i$, with a distribution

\begin{equation}
\label{p_W}
p_{\rm W}({\bf X})=\left<W_{\bf M} \left|
\delta(\hat{\bf X}-{\bf X}) \right| W_{\bf M}\right>.
\end{equation}

\noindent Substituting Eq. \ref{phi_wf} in Eq. \ref{precursor_mgf}, 
averaging over ${\bf k}$, and comparing with Eq. \ref{m_w}, after
discretizing the integral over ${\bf k}$, yields:

\begin{equation}
\label{mgf_wf}
C_{\rm W}({\boldsymbol \alpha}) = \frac{1}{N_c} \sum_{\bf k} 
{\cal C}({\bf k},{\boldsymbol \alpha}) = \left< W_{\bf M} \left|
e^{-\imath {\boldsymbol \alpha} \cdot \hat{\bf X}} \right| W_{\bf M} 
\right>,
\end{equation}

\noindent which shows that the function $C_{\rm W}({\boldsymbol \alpha})$ 
introduced in Section \ref{gi_cumulant} is 
the characteristic function of the distribution $p_{\rm W}({\bf X})$.
It is straightforward to check that the first moment of this distribution 
equals the first moment of the gauge-invariant polarization distribution 
$p({\bf X})$ generated by $C({\boldsymbol \alpha})$; this is similar to the 
independent-electron case, where the average polarization is given by the sum 
of the centers of charge of the gauge-dependent Wannier functions\cite{ksv}. 
Likewise, the ``many-body Wannier functions'' $W_{\bf M}$ are gauge-dependent,
and as a result so are the higher cumulants of $p_{\rm W}({\bf X})$, in 
particular the quadratric spread. Notice that the gauge-dependence of the 
$W_{\bf M}$ implies that in general they overlap with one 
another\cite{foot_overlap}, and only in certain gauges are they localized to 
the point of being essentially non-overlapping (provided that the 
system is insulating). This is different from the usual single-body Wannier 
functions, which in general remain overlapping in any gauge (even though they
are orthogonal); in the many-body case we are free to choose the volume $V$ 
large enough so that with a judicious choice of gauge the $W_{\bf M}$  become 
non-overlapping in the high-dimensional configuration space.

We are now in a position to provide the link between the present formalism
(and, therefore, the Berry phase theory of polarization) and Kohn's theory of 
localization\cite{kohn,kohn68}. A major result of Kohn's work is the conjecture
that a general many-body insulating 
wavefunction $\Psi_{{\bf k}=0}$ breaks up into a sum of non-overlapping parts 
$\Psi_{\bf M}$, localized in disconnected regions ${\cal R}_{\bf M}$ in 
configuration space (Eq. \ref{psi_insulator}).
Upon inspection, one finds that the present ``many-body Wannier functions'' 
are nothing other than Kohn's functions, $W_{\bf M} = \sqrt{N_c} \Psi_{\bf M}$ 
[compare Eq. \ref{phi_wf} with Eq. 6.1 of Ref. \cite{kohn}], except that Kohn 
only considered gauges where they are localized in such a way that the overlap
becomes exponentially small:

\begin{equation}
\label{loc_insulator}
\Psi_{\bf M}({\bf x}_1,...,{\bf x}_N) 
\Psi_{{\bf M}^{'}}({\bf x}_1,...,{\bf x}_N) 
\doteq 0 \hspace{0.2in}
\mbox{for  } {\bf M}^{'} \not= {\bf M},
\end{equation}

\noindent so that the $\Psi_{\bf M}$ are uniquely defined  
apart from exponentially small variations and a global phase. 
Transposing the language of Ref. \cite{marzari97} to a many-body framework, 
Kohn's functions $\Psi_{\bf M}$ can be viewed as the maximally-localized 
``many-body Wannier functions'' $W_{\bf M}$ 
(see Appendix \ref{omega_i_many_body}).

In such ``non-overlapping gauges'', the precusor characteristic function,  
Eq. \ref{precursor_mgf}, becomes, using  Eqs. \ref{phi_wf} and 
\ref{loc_insulator},

\begin{equation}
\label{precursor_mgf_localized}
{\cal C}({\bf k},{\boldsymbol \alpha}) \doteq 
N_c \left< \Psi_{\bf M}
\left| e^{-\imath {\boldsymbol \alpha} \cdot \hat{\bf X}} \right|
\Psi_{\bf M} \right> \mbox{  (``non-overlapping gauge'')},
\end{equation}

\noindent which is independent of ${\bf k}$, since the ${\bf k}$-dependence 
of ${\cal C}({\bf k},{\boldsymbol \alpha})$ in a general gauge arises from the
cross terms between different ${W}_{\bf M}$. Therefore in such gauges we 
recover Eq. \ref{clausius_mossotti}, which for a finite system obeying twisted
boundary conditions was valid in any gauge:

\begin{equation}
\label{clausius_mossotti_like}
{\cal C}({\bf k},{\boldsymbol \alpha}) \doteq 
C_{\rm W}({\boldsymbol \alpha}) \doteq 
C({\boldsymbol \alpha}) \mbox{  (``non-overlapping gauge'')}.
\end{equation}

\noindent Finally, from the gauge-invariance of $C({\boldsymbol \alpha})$ 
we conclude that {\it in any gauge}

\begin{equation}
\label{characteristic_kohn}
C({\boldsymbol \alpha}) \doteq 
N_c \left< \Psi_{\bf M} \left| e^{-\imath {\boldsymbol \alpha} \cdot 
\hat{\bf X}} \right| \Psi_{\bf M} \right>,
\end{equation}

\noindent which is the desired relation. It should be stressed that in deriving
this equation we {\it assumed} Kohn's conjecture regarding the existence of 
non-overlapping many-body Wannier functions in the insulating state.

We interpret Eq. \ref{characteristic_kohn} as follows: in each region 
${\cal R}_{\bf M}$ in the configuration space of the variables 
$\{ {\bf x}_i \}$ (Fig. 2), the variable ${\bf X} = \sum_{i=1}^N {\bf x}_i$ 
takes on a range of values with a distribution generated by 
$C({\boldsymbol \alpha})$,

\begin{equation}
\label{p_X_bulk}
p({\bf X}) \doteq N_c \left<\Psi_{\bf M} \left|
\delta(\hat{\bf X}-{\bf X}) \right| \Psi_{\bf M}\right>,
\end{equation}

\noindent which, together with Eq. \ref{moments_polarization_bulk}, gives the
bulk polarization distribution. The previous equation can therefore be viewed 
as the generalization for extended insulating systems of Eq. \ref{X_dist}, 
which gives the distribution of the electronic center of mass for finite 
systems. 

In Fig. 3 is represented the distribution $p(X_i)$ along the $i$-th direction.
According to Eq. \ref{loc_length},
the width of this distribution is $\sqrt{N} \xi_i(N)$.  
The solid lines describe what happens in a 1D insulator, for which 
the width of each of the peaks 
$p(X)$ labelled by $M$ is
$\sqrt{N} \xi(N) \propto \sqrt{L}$ and the distance between the centers of 
consecutive peaks is $L$, so that for large $L$ they 
are well-separated. In the case of a 3D insulator where all the linear 
dimensions are similar ($V \sim L^3$), the peaks $p(X_i)$ overlap for large 
$L$, since their width is $\sqrt{N} \xi_i(N) \propto L^{3/2}$ (dashed lines in
Fig. 3.) The important observation is that, according to 
Eqs. \ref{loc_insulator} and \ref{p_X_bulk}, even if they are 
{\it overlapping}, the distributions $p({\bf X})$ are well-defined 
``projections'' into real space of essentially {\it disconnected} distributions
${\left| \Psi_{\bf M} \right|}^2$ in configuration 
space\cite{foot_difference_kohn}.

At this point we shall reconsider the problem of how to define a meaningful
many-body position operator for the electrons in extended systems, an issue 
which has been recently discussed in Ref. \cite{resta98}. We recall that the 
source of the problem is that in such systems the usual center-of-mass operator
$\hat{\bf X}$ is ill-defined (see Section \ref{gen_extended}).
Kohn\cite{kohn,kohn68} proposed the following operator
as the substitute for $\hat{\bf X}$ for taking expectation values over the
ground state wavefunction of an extended {\it insulator}: 

\begin{equation}
\label{pos_kohn}
\hat{\bf Q} = \left\{ \begin{array}{ll}
\hat{\bf X} - {\bf R}_{\bf M} \mbox{ in } {\cal R}_{\bf M} \\
{\bf F}({\bf x}_1,...,{\bf x}_N) \mbox{ outside all } {\cal R}_{\bf M} \\
\end{array},
\right.
\end{equation}

\noindent where ${\bf F}$ is defined in a similar way as in 
Eq. \ref{pos_clausius}.
It is straightforward to check that
the moments calculated from $\hat{\bf Q}$ coincide with those derived from the
characteristic function: with the help of Eqs. \ref{psi_insulator}, 
\ref{characteristic_kohn}, and \ref{pos_kohn}, we find

\begin{eqnarray}
\label{moments_pos_kohn}
& \left< \Psi_{{\bf k}=0} \left| {\hat{Q}_j}^n 
\right| \Psi_{{\bf k}=0} \right> \doteq \nonumber \\
&N_c \left< \Psi_{{\bf M}= 0} \left| \hat{X}_j^n 
\right| \Psi_{{\bf M}= 0} \right> \doteq
\imath^n {\left. \partial^n_{\alpha_j^n} 
C({\boldsymbol \alpha}) \right|}_{{\boldsymbol \alpha}={\bf 0}}.
\end{eqnarray}

\noindent The connection with Ref. \cite{resta98} will be made in
Section \ref{discrete_fix}.

\subsection{Insensitivity of bulk properties to the 
boundary conditions}\label{insensitive_bc}

We now have an adequate framework for discussing the implications of 
localization for the dependence on boundary conditions of the bulk properties 
of insulators. As a first example, consider the adiabatic current 
${\bf J}_{\bf k}^{(\lambda)}$. Since it is gauge-invariant, we may evaluate it
in any gauge, in particular in a ``non-overlapping gauge''. Substituting 
Eq. \ref{phi_wf} into Eq. \ref{adiabatic_current} we find, because of the 
exponential decrease of the overlap as the size $L$ is increased: 

\begin{equation}
\label{adiabatic_current_b}
{\bf J}^{(\lambda)}_{\bf k} \doteq
q_e N_c \partial_{\lambda} \left< \Psi_{\bf M}^{(\lambda)} \left| 
\hat{\bf X} \right| \Psi_{\bf M}^{(\lambda)} \right> \doteq 
q_e \partial_{\lambda} {\left< {\bf X} \right>}^{(\lambda)},
\end{equation}

\noindent where for the last equality Eqs. \ref{cumulants_ext} and
\ref{characteristic_kohn} were used. Thus for a large system size 
${\bf J}_{\bf k}^{(\lambda)}$ is essentially 
${\bf k}$-independent (this was demonstrated in Ref. \cite{niu84} using a
different reasoning, for insulators with an energy gap).

As a second example, let us look at the quantum geometric tensor 
$T_{ij}({\bf k})$ and its real part, the metric tensor $G_{ij}({\bf k})$.
Evaluating Eq. \ref{geometric_tensor} in a ``non-overlapping gauge'' with the 
help of Eq. \ref{phi_wf}, we find 
$T_{ij}({\bf k}) \doteq
\left<  X_i X_j  \right> - \left< X_i \right> \left< X_j \right>$, 
which again is essentially independent of ${\bf k}$ for large systems (it is 
also real, so that $T_{ij}({\bf k}) \doteq G_{ij}({\bf k})$). Thus we conclude
that the ${\bf k}$-independent metric tensor gives the mean-square fluctuation 
of the polarization. This is a stronger statement than the one made at the end
of Section \ref{gi_cumulant}, which pertained to the ${\bf k}$-averaged metric
(by contrast, the single-electron metric tensor $g_{ij}({\bf k})$ given by 
Eq. \ref{metric} is in general ${\bf k}$-dependent, since the one-electron
Wannier functions remain overlapping even when they are maximally-localized; as
a consequence, the quadratic polarization fluctuations are related to its 
{\it average} over the Brillouin zone, as shown in 
Appendix \ref{omega_I_fluct}.)

These two examples 
illustrate an important point: the physically measurable ground state
quantities, such as
the polarization current and the polarization fluctuations, are well-defined
for any single boundary condition ${\bf k}$, 
and the choice of ${\bf k}$ becomes immaterial for large system sizes.
It may therefore seem bizarre that in the generating 
function formalism one needs to average over all ${\bf k}$ before obtaining 
gauge-invariant quantities, as shown in Section \ref{gi_cumulant}. 
A concrete example helps to clarify this state of
affairs: in the case of the second cumulant, had we taken the second derivative
of the {\it precursor} cumulant generating function,
$\ln {\cal C}({\bf k},{\boldsymbol \alpha})$, {\it without} afterwards 
averaging over ${\bf k}$, we would be left with a quantity (the integrand
of Eq. \ref{second_cumulant}), which is different from the desired one, 
$-T_{ij}({\bf k})$, and is gauge-dependent. However, the ${\bf k}$-average of 
that gauge-dependent quantity {\it is} gauge-invariant, and equals the 
${\bf k}$-average of $-T_{ij}({\bf k}) \doteq -G_{ij}({\bf k})$ 
(compare Eqs. \ref{second_cumulant} and 
\ref{second_cumulant_c}). A similar situation occurs with the adiabatic
polarization current.
 
The insensitivity of the bulk properties of insulators to the choice of
boundary conditions on a large system seems to be a very general property. 
Indeed, 
if $\hat{\cal O}$ is any well-defined operator acting on the ground-state of 
the extended system (e.g., the hamiltonian $\hat{H}$), then from 
Eqs. \ref{phi}, \ref{phi_wf}, and \ref{loc_insulator} it follows that
$\left< \Psi_{\bf k} \left| \hat{\cal O} \right| \Psi_{\bf k} \right> 
\doteq 
\left< \Psi_{{\bf k}=0} \left| \hat{\cal O} \right| \Psi_{{\bf k}=0} \right>$,
which helps to understand why, for instance, in insulators the Drude weight
(see Table I) goes to zero exponentially with the system size\cite{kohn}. This
has been confirmed by numerical simulations\cite{scalapino} and calculations on
exactly solvable models\cite{stafford91}. More precisely, in 
Ref. \cite{stafford91} it was found that the Drude weight for the half-filled 
Hubbard model in 1D scales as $\sim L^{1/2} e^{-L/\tilde{\xi}}$, which yields a
localization length $\tilde{\xi}$. It would be worthwhile to investigate
whether $\tilde{\xi}$ is the same as the localization length $\xi$ used in the
present work.

\section{Discretized formulas}\label{discrete}

Until now we have dealt with a continuum formalism, in which the
cumulants are obtained by differentiating $\ln C({\boldsymbol \alpha})$ at 
${\boldsymbol \alpha}=0$. Heuristically, this can be viewed as a measure of the
${\bf k}$-averaged change in $\Phi_{\bf k}$ as the boundary condition ${\bf k}$
is twisted adiabatically (see Eqs. \ref{precursor_mgf}, \ref{bulk_M}, and 
\ref{cumulants_ext}). In numerical calculations it is often more natural to 
perform {\it independent} calculations on a discrete mesh of ${\bf k}$-points 
and then use the resulting wavefunctions to estimate the derivatives by finite 
differences\cite{ksv,marzari97}. Since the global phases of the wavefunctions 
on the grid are unrelated, useful discretized expressions must remain 
invariant under arbitrary changes of those phases. The derivation of such
discretized formulas becomes particularly transparent in the present formalism,
as we will now show.

\subsection{Formulas involving an average over ${\bf k}$}
\label{discrete_avg}

We will find it convenient to work with scaled coordinates:
let the simulation cell be defined by the vectors 
$\{ {\bf L}_1, {\bf L}_2, {\bf L}_3 \}$, to which correspond the
reciprocal lattice vectors $\{ {\bf G}^1, {\bf G}^2, {\bf G}^3 \}$
(${\bf L}_i \cdot {\bf G}^j = 2 \pi \delta_i^{\hspace{0.02in} j}$). The scaled
coordinates $S^i$ are defined by ${\bf X} = S^i {\bf L}_i$, where a sum over 
repeated indices is implied, and similarly we have ${\bf k} = q_i {\bf G}^i$ 
and ${\boldsymbol \alpha} = t_i {\bf G}^i$ (and thus 
${\boldsymbol \alpha} \cdot {\bf X} = 2 \pi t_i S^i$). In terms of these
variables, Eqs. \ref{precursor_mgf} and \ref{bulk_M} become

\begin{equation}
\label{precursor_mgf_scaled}
{\cal C}({\bf q},{\bf t}) = 
\left<\Psi_{\bf q} \left| e^{-\imath 2 \pi t_i S^i}  
\right| \Psi_{{\bf q}+{\bf t}} \right> =
\left< \Phi_{\bf q} \left| \Phi_{{\bf q}+{\bf t}} \right.
\right>,
\end{equation}

\noindent and

\begin{equation}
\label{bulk_M_scaled}
\ln C({\bf t}) =
\int_0^1 dq_1 \int_0^1 dq_2 \int_0^1 dq_3 \ln {\cal C}({\bf q},{\bf t}),
\end{equation}

\noindent so that $C({\bf t})$ is the joint characteristic function
for the variables $2 \pi S^i$.

The average is

\begin{equation}
\label{first_moment_scaled}
{\left< S^l \right>} = \frac{\imath}{2 \pi} \int d{\bf q}
{\left. \partial_{t_l} \ln {\cal C}({\bf q},{\bf t}) \right|}_{{\bf t}=0}.
\end{equation}

\noindent Next we discretize the derivative in the integrand as

\begin{equation}
\label{first_derivative_discretized}
\delta q_l {\left.
\partial_{t_l} \ln {\cal C}({\bf q},{\bf t}) \right|}_{{\bf t}=0} \simeq
\ln {\cal C}({\bf q},\delta {\bf q}_l) - \ln {\cal C}({\bf q},0)
\simeq \imath {\rm Im} \ln {\cal C}({\bf q},\delta {\bf q}_l),
\end{equation}

\noindent where we made use of Eq. \ref{precursor_mgf}. This expression is 
gauge-dependent, just like its continuum counterpart, the Berry connection. 
As in the continuum case, gauge-invariance is recovered by averaging over
${\bf k}$;
that is done by choosing a uniform row of $J+1$ ${\bf k}$-points along the 
direction of ${\bf G}^l$ such that the endpoints are separated by ${\bf G}^l$,
${\bf k}_{\gamma} = {\bf k}_{\perp}+(\gamma/J){\bf G}^l$ with $\gamma=0,...,J$;
then $\delta q_l =1/J$, and we find:

\begin{equation}
\label{first_moment_scaled_b}
\left< S^l \right> \simeq - \frac{1}{2 \pi} \int
dq_i dq_j \sum_{\gamma=0}^{J-1} {\rm Im} 
\ln \left< \Phi_{{\bf k}_{\gamma}} \left| \Phi_{{\bf k}_{\gamma+1}} \right.
\right>,
\end{equation}

\noindent where $(i,j,l)$ is some permutation of $(1,2,3)$, and the periodic 
gauge is enforced by setting
$\Phi_{{\bf k}_J} = e^{-\imath {\bf G}^l \cdot {\bf X}} 
\Phi_{{\bf k}_0}$. The average polarization is then given by

\begin{equation}
\label{polarization_discrete}
\left< {\bf P}_{\rm el} \right> \cdot {\bf G}^l = \frac{2 \pi q_e}{V}
{\left< S^l \right>} \simeq -\frac{q_e}{V} \int dq_i dq_j
{\rm Im} \ln  \prod_{\gamma=0}^{J-1}
\left< \Phi_{{\bf k}_{\gamma}} \left| \Phi_{{\bf k}_{\gamma+1}} \right. 
\right>,
\end{equation}

\noindent which is the many-body analogue of the discretized 
Berry phase formula proposed in Ref. \cite{ksv}. It is straightforward to 
verify that it is gauge-invariant. This property hinges upon (i) using a
uniform row of ${\bf k}$-points, (ii) choosing a periodic gauge, and 
(iii) having a logarithm in the expression, which in the present derivation 
appears quite naturally, coming from the cumulant generating function, 
Eq. \ref{bulk_M_scaled}\cite{foot_log}.

To compute the variance we discretize the second logarithmic derivative:

\begin{equation}
\label{second_der_discretized}
{(\delta q_l)}^2 {\left.
\partial_{t_l^2}^2 \ln {\cal C}({\bf q},{\bf t}) \right|}_{{\bf t}=0} \simeq
\ln {\cal C}({\bf q},\delta {\bf q}_l) 
+ \ln {\cal C}({\bf q},-\delta {\bf q}_l)
- 2 \ln {\cal C}({\bf q},0) = 
\ln \left[ {\cal C}({\bf q},\delta {\bf q}_l)
{\cal C}({\bf q},-\delta {\bf q}_l) \right],
\end{equation}

\noindent which is gauge-dependent, similarly to 
Eq. \ref{first_derivative_discretized}\cite{foot_even_derivatives}. 
After some manipulations we obtain the following gauge-invariant, 
${\bf k}$-averaged formula:

\begin{equation}
\label{variance_scaled}
\left< {\left( S^l \right)}^2 \right>_c \simeq - \frac{J}{{(2\pi)}^2}
\int dq_i dq_j \ln {\left| \prod_{\gamma=0}^{J-1} 
\left< \Phi_{{\bf k}_{\gamma}} \big| \Phi_{{\bf k}_{\gamma+1}} \right> 
\right|}^2. 
\end{equation}

The evaluation of the covariance requires taking cross derivatives, which
can be discretized as

\begin{eqnarray}
\label{mixed_not_stirred}
&\partial^2_{xy} f \simeq
\frac{1}{2 \delta x \delta y} \bigl[
 f(x+\delta x,y+\delta y)+f(x-\delta x,y-\delta y) \nonumber\\
&-f(x+\delta x,y)-f(x-\delta x,y)-f(x,y+\delta y)-f(x,y-\delta y)+2f(x,y) 
\bigr],
\end{eqnarray}

\noindent leading to

\begin{eqnarray}
\label{covariance_scaled}
&{\left< S^j S^l \right>}_c \simeq - \frac{1}{2{(2 \pi)}^2} 
\int d q_i \Bigl[ \ln \prod_{\nu = 0}^{J_j-1}
\prod_{\gamma = 0}^{J_l-1}
{\left| \left< \Phi_{{\bf k}_{\nu \gamma}} \left|
\Phi_{{\bf k}_{\nu+1,\gamma+1}} \right. \right>\right|}^2 \nonumber \\
&-\ln \prod_{\nu=0}^{J_j-1}  \prod_{\gamma=0}^{J_l-1}
{\left| \left< \Phi_{{\bf k}_{\nu \gamma}} \left|
\Phi_{{\bf k}_{\nu+1,\gamma}} \right. \right> \right|}^2 -
\ln \prod_{\nu=0}^{J_j-1}  \prod_{\gamma=0}^{J_l-1}
{\left| \left< \Phi_{{\bf k}_{\nu \gamma}} \left|
\Phi_{{\bf k}_{\nu,\gamma+1}} \right. \right> \right|}^2
\Bigr],
\end{eqnarray}

\noindent where ${\bf k}_{\nu\gamma} = 
{\bf k}_{\perp} + \frac{\nu}{J_j}{\bf G}^j + \frac{\gamma}{J_l}{\bf G}^l$.

In order to convert back to cartesian coordinates we use the transformation law
for a second-rank contravariant tensor,
${\left< X_i X_j \right>}_c = H_{il} {\left< S^l S^m \right>}_c H_{jm}$,
where $H = \{{\bf L}_1,{\bf L}_2,{\bf L}_3\}$.
Eqs. \ref{polarization_discrete}, \ref{variance_scaled}, and
\ref{covariance_scaled}
are all that is needed
to calculate the average polarization and its quadratic fluctuations, as well 
as the correlations along different cartesian directions (if the cell symmetry
is orthorhombic or higher, Eq. \ref{covariance_scaled} is not needed for
computing the fluctuations.) In Appendix \ref{frac_fill} we give the required
modifications to deal with fractional filling.

\subsection{Formulas involving a single ${\bf k}$}\label{discrete_fix}

We now turn to a different kind of discretized formulas, the so-called
``single-point'' formulas, such as Eq. \ref{z_N}. In the case of localization
they were previously given for 1D only\cite{resta99}. In the present formalism
the generalization to higher dimensions becomes straightforward.
We will arrive at the ``single-point'' formulas 
starting from the expressions
derived in the previous Section, which involve averages over a uniform grid of
${\bf k}$-points. The basic idea is to perform the many-body analogue of a 
``Brillouin zone folding''\cite{resta_lect}. Let us start by 
discretizing the remaining integrals 
in Eqs. \ref{polarization_discrete}, \ref{variance_scaled}, and
\ref{covariance_scaled}; the expression for the average polarization, for
example, becomes

\begin{equation}
\label{polarization_intermediate} 
\left< {\bf P}_{\rm el} \right> \cdot {\bf G}^l \simeq 
- \frac{q_e}{V} \frac{1}{J_i J_j} {\rm Im} \ln
\prod_{\mu=0}^{J_i-1} \prod_{\nu=0}^{J_j-1} \prod_{\gamma=0}^{J_l-1}
\left< \Phi_{{\bf k}_{\mu \nu \gamma}}\left| 
\Phi_{{\bf k}_{\mu \nu, \gamma+1}} \right. \right>,
\end{equation}

\noindent where we have used a 
uniform spacing along all the reciprocal lattice directions:
${\bf k}_{\mu\nu\gamma} = \tilde{\bf k} + (\mu/J_i){\bf G}^i +
(\nu/J_j){\bf G}^j + (\gamma/J_l){\bf G}^l$ and $\tilde{\bf k}$ is
fixed (usually $\tilde{\bf k}=0$.)
Next we build an {\it ansatz} wavefunction\cite{resta99} 
$\tilde{\Psi}_{\tilde{\bf k}}$ 
contaning $\tilde{N} = JN$ electrons as the antisymmetrized product of the
$J=J_1J_2J_3$ separate $N$-electron wavefunctions 
$\Psi_{{\bf k}_{\mu \nu \gamma}}$:

\begin{equation}
\label{antisymmetrized}
\tilde{\Psi}_{\tilde{\bf k}} = \frac{1}{\sqrt{\tilde{N}!{N!}^{J}}}
\sum_{P}{(-1)}^P P \Psi(1,...,\tilde{N}), 
\end{equation}

\noindent where the sum is over all the permutations $P$ of 
the $\tilde{N}$ particle coordinates, ${(-1)}^P$ is the parity of the 
permutation, and
$\Psi(1,...,\tilde{N}) = \prod_{\mu \nu \gamma} 
\Psi_{{\bf k}_{\mu \nu \gamma}}$.
It is easy to verify that $\tilde{\Psi}_{\tilde{\bf k}}$ obeys
$\tilde{\bf k}$ boundary conditions on the larger cell containing $\tilde{N}$
electrons. It can be shown that for even $N$ 
this leads to (compare with Eq. 16 in Ref. \cite{resta99}, which deals with the
1D case):

\begin{equation}
\label{bz_folding_1}
\prod_{\mu\nu\gamma} \left< \Phi_{{\bf k}_{\mu \nu \gamma}} \left| 
\Phi_{{\bf k}_{\mu \nu, \gamma+1}} \right. \right> = 
\left< \tilde{\Psi}_{\tilde{\bf k}} \left| 
e^{- \imath \tilde{\bf G}^l \cdot \hat{\tilde{\bf X}}} \right|
\tilde{\Psi}_{\tilde{\bf k}} \right>,
\end{equation}

\noindent where $\tilde{\bf G}^l = {\bf G}^l / J_l$ is a basis vector of the
reciprocal of the larger cell, and
$\hat{\tilde{\bf X}}$ is the center of mass position operator of the
$\tilde{N}$-electron system. 
Comparison with Eq. \ref{polarization_intermediate} yields

\begin{equation}
\label{polarization_single_pt} 
\left< {\bf P}_{\rm el} \right> \cdot \tilde{{\bf G}}^l \simeq 
- \frac{q_e}{\tilde{V}} {\rm Im} \ln
\left< \tilde{\Psi}_{\tilde{\bf k}} \left| 
e^{- \imath \tilde{\bf G}^l \cdot \hat{\tilde{\bf X}}} \right|
\tilde{\Psi}_{\tilde{\bf k}} \right>.
\end{equation}

Similarly, discretizing the integrals in Eq. \ref{variance_scaled} 
and using Eq. \ref{bz_folding_1}, we find

\begin{equation}
\label{variance_single_pt}
{\left< {\left( S^l \right)}^2 \right>}_c \simeq - \frac{1}{{(2\pi)}^2}
\frac{J_l}{J_i J_j} \ln 
{\left| \left< \tilde{\Psi}_{\tilde{\bf k}} \left|
 e^{- \imath \tilde{\bf G}^l \cdot \hat{\tilde{\bf X}}} \right|
 \tilde{\Psi}_{\tilde{\bf k}} \right> 
\right|}^2.
\end{equation}

\noindent The quantity on the l.h.s. pertains to the $N$-particle system, 
whereas those on the r.h.s. pertain to the $\tilde{N}$-particle system. 
However, if we choose $J_1 = J_2 = J_3$ and use
$\left< X_i X_j \right>_c /N \simeq 
\left< \tilde{X}_i \tilde{X}_j \right>_c /\tilde{N}$ together with the
transformation law between cartesian and scaled coordinates, 
we find
$\left< \tilde{S}^i \tilde{S}^j \right>_c =
J_1 \left< S^i S^j \right>_c$. Eq. \ref{variance_single_pt} then becomes

\begin{equation}
\label{variance_single_pt_b}
{\left< {\left( \tilde{S}^l \right)}^2 \right>}_c \simeq - \frac{1}{{(2\pi)}^2}
\ln {\left| \left< \tilde{\Psi}_{\tilde{\bf k}} \left|
 e^{- \imath \tilde{\bf G}^l \cdot \hat{\tilde{\bf X}}} \right|
 \tilde{\Psi}_{\tilde{\bf k}} \right> 
\right|}^2,
\end{equation}

\noindent where all the quantities are now explicitly written for the
$\tilde{N}$-particle system.

The following relation can be derived along the same lines as 
Eq. \ref{bz_folding_1}:

\begin{equation}
\label{bz_folding_2}
\prod_{\mu\nu\gamma} \left< \Phi_{{\bf k}_{\mu \nu \gamma}} \left| 
\Phi_{{\bf k}_{\mu \nu+1, \gamma+1}} \right. \right> = 
\left< \tilde{\Psi}_{\tilde{\bf k}} \left| 
e^{- \imath \tilde{\bf G}^j \cdot \hat{\tilde{\bf X}}}
e^{- \imath \tilde{\bf G}^l \cdot \hat{\tilde{\bf X}}} \right|
\tilde{\Psi}_{\tilde{\bf k}} \right>.
\end{equation}

\noindent Discretizing the remaining integral in Eq. \ref{covariance_scaled} 
and again setting $J_1=J_2=J_3$, we obtain, using Eqs. \ref{bz_folding_1} and
\ref{bz_folding_2},

\begin{equation}
\label{covariance_single_pt}
\left< \tilde{S}^j \tilde{S}^l \right>_c \simeq - \frac{1}{2{(2 \pi)}^2}
\left[ \ln {\left| \left< \tilde{\Psi}_{\tilde{\bf k}} \left|
e^{- \imath \tilde{\bf G}^j \cdot \hat{\tilde{\bf X}}}
e^{- \imath \tilde{\bf G}^l \cdot \hat{\tilde{\bf X}}} \right|
\tilde{\Psi}_{\tilde{\bf k}} \right> \right|}^2 -
\ln {\left| \left< \tilde{\Psi}_{\tilde{\bf k}} \left|
e^{- \imath \tilde{\bf G}^j \cdot \hat{\tilde{\bf X}}}
\right| \tilde{\Psi}_{\tilde{\bf k}} \right> \right|}^2 -
\ln {\left| \left< \tilde{\Psi}_{\tilde{\bf k}} \left|
e^{- \imath \tilde{\bf G}^l \cdot \hat{\tilde{\bf X}}}
\right| \tilde{\Psi}_{\tilde{\bf k}} \right> \right|}^2
\right].
\end{equation}
  
Eqs. \ref{polarization_single_pt}, \ref{variance_single_pt_b}, and 
\ref{covariance_single_pt} are the desired ``single-point'' formulas.
In Appendix \ref{frac_fill} we modify them for the case of fractional filling.

In 1D the above formulas become particularly simple, and we recover the results
of Refs. \cite{resta98} and \cite{resta99}. In cartesian coordinates
($\tilde{X}=\tilde{S}\tilde{L}$ and $\tilde{G} = 2 \pi / \tilde{L}$) they are

\begin{equation}
\label{avg_dscrt_sngl_pt}
\left< \tilde{X} \right> \simeq - \frac{\tilde L}{2 \pi} {\rm Im} \ln
\left< \tilde{\Psi}_{\tilde{k}} \left| 
e^{- \imath \frac{2 \pi}{\tilde L} \hat{\tilde{X}}} \right|
\tilde{\Psi}_{\tilde{k}} \right>,
\end{equation}

\noindent which is equivalent to Eq. \ref{P_resta}, and

\begin{equation}
\label{variance_sngl_pt}
\left< {\Delta \tilde{X}}^2 \right> \simeq - \frac{\tilde{L}^2}{{(2\pi)}^2}
\ln {\left| \left< \tilde{\Psi}_{\tilde{k}} \left| 
e^{- \imath \frac{2 \pi}{\tilde L} \hat{\tilde{X}}} \right|
\tilde{\Psi}_{\tilde{k}} \right> \right|}^2,
\end{equation}

\noindent which, together with Eq. \ref{loc_length}, gives 
Eq. \ref{resta_lambda}.

\subsection{Interpretation of the ``single-point'' formulas }
\label{comparison}

In the previous Section we derived ``single-point'' formulas for the first two
moments of the polarization,
starting from discrertized expressions involving an average over boundary 
conditions.
In order to switch between the two descriptions, an {\it ansatz} wavefunction 
${\tilde\Psi}_{\tilde k}$ given by Eq. \ref{antisymmetrized} was used. However,
this is strictly valid only if the $\tilde{N}$ particles are not correlated 
with each other. In that case ${\tilde\Psi}_{\tilde k}$ becomes a Slater 
determinant of one-electron orbitals, and the procedure leading to the 
``single-point'' formulas (``Brillouin zone folding'') does not involve any
further approximations (see Refs. \cite{resta_lect} and \cite{resta98}). 

For a correlated state of
many particles the situation is rather different. In that context
the ``single-point'' formulas were originally proposed in 1D
for an {\it arbitrary}
correlated $\tilde{N}$-electron wavefunction ${\tilde\Psi}_{{\tilde k}=0}$ with
periodic boundary conditions over a cell of size 
$\tilde{L}$\cite{resta98,resta99}; such wavefunction in general will not obey 
Eq. \ref{antisymmetrized}, of course. The derivation of
the previous Section allows us to assess the approximations involved:
these have to do with the extent to which a wavefunction
given by Eq. \ref{antisymmetrized} 
differs from the fully correlated wavefunction. The key quantity to consider is
the correlation length $L_{\rm corr}$, which quantifies the range over which 
the particles are correlated (in addition to any long range order which is 
included in the mean field potentials seen by each particle). The fact that the
correlations are short-ranged has been termed ``nearsightedness'' by Kohn in a
recent paper\cite{kohn96}; in fact, in an insulator, for example, one expects 
the longest range correlations to be of the Van der Waals type, which decay as
$1/r^6$ in the energy and as $1/r^3$ in the wavefunction\cite{fulde,mo97}. The
basic assumption underlying Eq. \ref{antisymmetrized}, namely that the 
$\tilde{N}$ electrons only correlate in groups of $N$ at a time\cite{resta99},
is consistent with the principle of nearsightedness. This is the justification
for applying the ``single-point'' formulas to a correlated insulating 
wavefunction, provided that the cell is large enough. 

From this perspective it becomes clear that the two types of formulas derived 
in Sections \ref{discrete_avg} and \ref{discrete_fix} constitute different 
approximations to the same continuum expressions, which involve an average over
all twisted boundary conditions. In both approaches one must choose cells with
linear dimensions greater than $L_{\rm corr}$: in the latter approach one must 
use a single large cell with size $JV$ in order to have the same level of
accuracy as $J$ independent calculations each of size $V$ using the former 
approach.

From a practical point of view, there are two possible ways to proceed:
either perform several calculations with different twisted boundary conditions
on a smaller cell and then use the formulas of Section \ref{discrete_avg}, or 
perform a single calculation on a large cell and then use the formulas of
Section \ref{discrete_fix}. Although the two approaches are comparable in terms
of accuracy, the former approach should be more efficient, because the 
available many-body algorithms do not scale linearly with the number of 
particles.
On the other hand, the averaging over boundary conditions may be more 
cumbersome to implement in practice.

\section{conclusions}\label{discussion}

We have presented in this work a unified theory of electronic polarization and
localization in insulators. The central quantity in
the formalism is the cumulant generating function $\ln C({\boldsymbol \alpha})$
defined in Eq. \ref{bulk_M}; it provides a systematic procedure for extracting
from the ground state wavefunction $\Psi_{\bf k}$ the moments of a properly
defined distribution $p({\bf X})$ for the electronic center of mass in an 
extended insulator. In complete analogy with the case of a finite system, this
distribution is simply related to the quantum distribution of the ground state
polarization, via Eq. \ref{moments_polarization_bulk}.

Several seemingly disparate ideas regarding electronic polarization and
localization in 
insulators\cite{kohn,resta99,kudinov} are brought together quite naturally 
using the generating function approach. In particular, it shows the connection
between 
the Berry phase theory of polarization\cite{ksv,om94}, as well
as the related theory of localization\cite{marzari97,resta99,aligia99b}, and 
Kohn's theory of localization in the insulating state\cite{kohn,kohn68}.
A key quantity is the localization length $\xi_i$, which is defined in terms of
the experimentally measurable mean-square fluctuation of the polarization 
(Eq. \ref{loc_length} and Table I), making contact with the work of 
Kudinov\cite{kudinov}. In the thermodynamic limit $\xi_i$ agrees with the 
localization length defined in Ref. \cite{resta99} for 1D systems
(Eq. \ref{resta_lambda}).
Furthermore, the generating function formalism also reveals a very close formal
analogy between Kohn's localized many-electron functions $\Psi_{\bf M}$ and the
maximally-localized one-electron Wannier functions defined by Marzari and
Vanderbilt\cite{marzari97}: the 
former can be viewed as maximally-localized ``many-body Wannier functions'' 
(see Section \ref{loc_config_space} and Appendix \ref{omega_i_many_body}). 
Moreover, in the same way that the quadratic spread of the functions 
$\Psi_{\bf M}$ is a measure of the mean-square fluctuations of the bulk 
polarization, the gauge-invariant part of the spread of the Wannier functions 
(which in 1D systems equals the spread of the maximally-localized Wannier
functions) measures the same quantity for uncorrelated insulators.
The fluctuation-dissipation relation can be used to derive an inequality
(Eq. \ref{inequality})
between the polarization fluctuations and the minimum energy gap for optical
absorption in an insulator. 
The present approach also provides some extra insight into the appearence of a 
``quantum of polarization'' in periodic insulators\cite{ksv}, which is seen to 
be related to 
the existence of multiple geometrically equivalent regions ${\cal R}_{\bf M}$,
depicted in Fig. 2 (see Appendix \ref{quantum}).

The localization length seems to play a role in the 
theory of insulators similar to that of the Drude weight in the theory of ideal
conductors: the latter measures how ``free'' the ``free charges'' in a perfect
conductor are, whereas the former measures how localized the ``bound charges''
in an insulator are. Interestingly, both quantities can be expressed as second
derivatives with respect to the twisted boundary conditions (see Table I),
which play a crucial role in the formalism.
As discussed in Section \ref{insensitive_bc}, {\it bulk properties} of 
insulators are rather insensitive to the boundary conditions, unlike the 
properties of conductors. Nevertheless, the insulating {\it wavefunction} 
itself is not insensitive to the boundary conditions\cite{niu84}; on the 
contrary, the derivatives of $\ln C({\boldsymbol \alpha})$, which measure the 
${\bf k}$-averaged change with ${\bf k}$ in 
$\Phi_{\bf k}=e^{-\imath {\bf k} \cdot {\bf X}} \Psi_{\bf k}$ as the boundary 
condition on
$\Psi_{\bf k}$ is twisted, contain quantitative information about basic 
properties of the insulator: in particular, the first derivative gives the 
average macroscopic polarization, and the second derivative gives its 
mean-square fluctuation. Both quantities have a geometrical
interpretation: the former is a Berry phase on a
manifold of quantum states parametrized by the twisted boundary 
conditions\cite{ksv,om94} (Eq. \ref{first_cumulant}), 
and the latter is a metric 
on the same manifold (Eq. \ref{second_cumulant_c}).

The generating function approach also leads naturally to discretized formulas
which can be used to compute the 
polarization and the localization in many-body numerical calculations in any 
number of dimensions. Two alternative kinds of expressions exist: those 
involving wavefunctions computed on a uniform grid of ${\bf k}$-points 
(Eqs. \ref{polarization_discrete}, \ref{variance_scaled}, and
\ref{covariance_scaled}), and those involving a single wavefunction with a 
fixed boundary condition $\tilde{\bf k}$ (Eqs. \ref{polarization_single_pt}, 
\ref{variance_single_pt_b}, and \ref{covariance_single_pt}).
The present derivation clearly shows how the two types of formulas are related
to one another.

\section*{ACKNOWLEDGMENTS}

We wish to thank David Vanderbilt for illuminating comments and 
suggestions for improvement on an early draft of the paper; Yu. Kagan,
Jos\'e Lu\'\i s Martins, Nicola Marzari,
Daniel S\'anchez-Portal, and John Shumway for stimulating discussions; 
J. L. M. for bringing to our attention Ref. \cite{hirst},
Gerardo Ortiz for Refs. \cite{kudinov} and
\cite{scalapino}, and Michael Stone for Ref. \cite{berry89}. 
This work was supported by the DOE Grant No. DEFG02-96-ER45439 and the
NSF Grant No. DMR9802373.
I.S. acknowledges financial support from FCT (Portugal).

\appendix

\section{Localization length and width of the Wannier functions}
\label{omega_I_fluct}

In Section \ref{loc_config_space} it was shown that the width of the electronic
center of mass distribution $p(X_i)$, arising from the maximally-localized 
``many-body Wannier function'' $\Psi_{\bf M}$, is $\sqrt{N} \xi_i$. In this
Appendix we work in the independent-electron framework, in which the usual
one-electron Wannier functions are defined, and investigate the relation
between their width and the localization length $\xi_i$ (i.e., the mean-square
fluctuation of the polarization).
It is clear that since the
spread (Eq. \ref{omega}) is gauge-dependent, it cannot relate directly to any 
measurable quantity. One might have guessed that, as happens with the 
``many-body Wannier functions'', the physically meaningful quantity would be 
the spread of the maximally-localized Wannier functions. Building upon the 
results of Ref. \cite{marzari97}, we show here that 
the gauge-invariant part $\Omega_{\rm I}$ of the spread of the occupied Wannier
functions (Eq. \ref{omega_I}) measures the mean-square fluctuation of the bulk
polarization. 

The proof follows from the fluctuation-dissipation theorem: for a crystalline
insulator in the 
independent-electron approximation, the real part of the optical conductivity 
due to interband vertical transitions is given by\cite{bassani}

\begin{equation}
\label{kubo_indep}
{\rm Re} \sigma_{ij}(\omega) =
\frac{\pi f q_e^2}{m_e^2 \hbar \omega} \sum_{n=1}^M \sum_{m = M+1}^{\infty}
\int_{\rm BZ} \frac{d {\bf k}}{{(2 \pi)}^3} 
      p_{nm,{\bf k}}^i p_{mn,{\bf k}}^j \delta(\omega_{mn}^{\bf k} - \omega), 
\end{equation}

\noindent where $f$ is the occupation number of states in the valence band (in
spin-degenerate systems $f=2$),
${\bf p}_{nm,{\bf k}} = - \imath \hbar \left< \psi_{n \bf k} | 
{\boldsymbol \nabla} |
\psi_{m \bf k} \right>$, and 
$\hbar \omega_{mn}^{\bf k} = \epsilon_{m \bf k} - \epsilon_{n \bf k}$ is the
difference between single-particle energies. Instead of Eq. \ref{commutation} 
we now have ${\bf p}_{nm,\bf k} = m_e \omega_{mn}^{\bf k} \left< u_{n \bf k} | 
{\partial}_{\bf k} u_{m \bf k} \right>$, and following similar steps
as in Section \ref{fdt} for the many-body case we find, using Eq. \ref{metric},

\begin{equation}
\label{metric_2}
\frac{\hbar}{\pi f q_e^2} \int_0^{\infty} \frac{d \omega}{\omega}
{\rm Re} \sigma_{ij}(\omega)=
\frac{1}{{(2 \pi)}^3} \int_{\rm BZ} d {\bf k} g_{ij}({\bf k}).
\end{equation}

\noindent Combining with Eqs. \ref{omega_I} and \ref{fdt_bulk}, using 
$v/(fV) = M/N$, and taking the trace, we obtain the fluctuation-dissipation
relation\cite{foot_composite}:

\begin{equation}
\label{omega_I_fdt}
\Omega_{\rm I} = \frac{\hbar v}{\pi f q_e^2} 
\int_0^{\infty} \frac{d \omega}{\omega} {\rm Tr}{\rm Re} \sigma(\omega) =
M \lim_{N \rightarrow \infty} \frac{\left< {\Delta {\bf X}}^2 \right>}{N}.
\end{equation}

\noindent Comparison with Eq. \ref{loc_length} shows that
$\Omega_{\rm I}/M = \sum_{i=1}^3 \xi_i^2$, which is the multidimensional
generalization of a result obtained in Ref. \cite{resta99}. 
It has been found that typically $\Omega_{\rm I}$ accounts for most of the 
spread $\Omega$; for instance, in semiconductors it usually accounts for more 
than 90\%\cite{marzari97}. 
Thus it is justifiable to view $\xi_i$ as an estimate of the average width 
along the $i$-th direction of the occupied Wannier functions (in 1D $\xi_i^2$ 
is actually the average spread of the maximally-localized 
ones\cite{marzari97}). This conclusion agrees with the analysis in Appendix 1 
of Ref. \cite{kohn}: there, the distribution $p(X)$ of the electronic center of
mass of a 1D non-interacting insulating system was studied, with the result 
that $X$ is localized in an interval of width $\sqrt{N} b$, where ``$b$ is a 
length of the order of the range of the Wannier function''.
It is clear from the results of Section \ref{loc_config_space} and the previous
discussion that the gauge-invariant $\xi$ is precisely that length $b$. 

It is straightforward to check that
for non-interacting crystalline insulators, Eq. \ref{inequality} becomes, after
summing over all $d$ cartesian directions,
$\Omega_{\rm I}/M \leq d \hbar^2/(2 m_e E_{\rm g})$, where $E_{\rm g}$ is the
minimum direct gap over the Brillouin zone. Hence in general the inequality
involves the gauge-invariant part of the spread. Since in 1D this equals the
average spread of the maximally-localized Wannier functions, we recover 
Kivelson's original result\cite{foot_kivelson}.

\section{spread of the ``many-body wannier functions''}
\label{omega_i_many_body}

In Section \ref{loc_config_space} we introduced the functions $W_{\bf M}$,
which play a role in the many-body theory of polarization similar to that of 
the Wannier functions in the independent-electron theory. Here we will show 
that this formal analogy carries over to considerations about the
spread of those functions. 
As mentioned in Section \ref{recent_develop}, in the independent-electron 
framework the quadratic spread $\Omega$ of the occupied Wannier functions can 
be decomposed into a sum of two positive terms, one of them ($\Omega_{\rm I}$)
gauge-invariant\cite{marzari97}; hence $\Omega_{\rm I} \leq \Omega$ in any 
gauge. Following similar steps as in Ref. \cite{marzari97}, here we will derive
the corresponding result for the many-body functions $W_{\bf M}$. First we 
define their gauge-dependent spread by analogy with Eq. \ref{omega}:

\begin{equation}
\label{omega_many_body}
\Omega^{({\rm W})}=
\left< W_{\bf 0} \left| {\hat{\bf X}}^2 \right| W_{\bf 0} 
\right>-
{\left< W_{\bf 0} \left| \hat{\bf X} \right| W_{\bf 0} \right>}^2 =
- {\left. \partial^2_{{\boldsymbol \alpha}^2}
\ln C_{\rm W}({\boldsymbol \alpha}) \right|}_{{\boldsymbol \alpha}=0},
\end{equation}

\noindent where $C_{\rm W}({\boldsymbol \alpha})$ is given by Eq. 
\ref{mgf_wf}. Next we show that the role of $\Omega_{\rm I}$ is played by 
$\left< {\Delta {\bf X}}^2 \right> = - {\left. 
\partial^2_{{\boldsymbol \alpha}^2}
\ln C({\boldsymbol \alpha}) \right|}_{{\boldsymbol \alpha}=0}$:
taking this derivative inside the integral
in Eq. \ref{bulk_M}, discretizing that integral\cite{foot_continuum},
and comparing with Eq. \ref{mgf_wf}, this becomes

\begin{equation}
\label{omega_I_many_body_b}
\left< {\Delta {\bf X}}^2 \right> = 
\left< W_{\bf 0} \left| {\hat{\bf X}}^2 \right| W_{\bf 0} \right> -
\frac{1}{N_c} \sum_{\bf k} {\left(\imath \left< \Phi_{\bf k} \big|
\partial_{\bf k} \Phi_{\bf k} \right> \right)}^2.
\end{equation}

\noindent Now we use the relation 

\begin{equation}
\label{berry_connection}
\imath \left< \Phi_{\bf k} \big| \partial_{\bf k} \Phi_{\bf k} \right> =
\sum_{\bf M} e^{-\imath {\bf k} \cdot {\bf R}_{\bf M}}
\left< W_{\bf M} \left| \hat{\bf X} \right| W_{\bf 0} \right>,
\end{equation}

\noindent which is the many-body analogue of Eq. 5 of Ref. \cite{marzari97}
and can be derived in the same way. Substituting into 
Eq. \ref{omega_I_many_body_b} gives

\begin{equation}
\label{omega_I_many_body_c}
\left< {\Delta {\bf X}}^2 \right> = 
\left< W_{\bf 0} \left| {\hat{\bf X}}^2 \right| W_{\bf 0} \right> -
\sum_{\bf M} {\left| 
\left< W_{\bf M} \left| \hat{\bf X} \right| W_{\bf 0} \right> 
\right|}^2,
\end{equation}

\noindent and comparison with Eq. \ref{omega_many_body} yields the desired 
result: $\left< {\Delta {\bf X}}^2 \right> \leq \Omega^{({\rm W})}$. 

There is, however, an important difference with respect to the single-electron
Wannier functions: in Ref. \cite{marzari97} it was shown that it is 
only in 1D that the maximally-localized one-electron Wannier functions have a 
spread $\min [\Omega]=\Omega_{\rm I}$, whereas in higher dimensions 
$\min [\Omega] > \Omega_{\rm I}$, i.e., $\Omega_{\rm I}$ is strictly a lower 
bound to $\Omega$. As discussed in Section \ref{loc_config_space}, according to
Kohn\cite{kohn,kohn68} a gauge can be chosen where the functions 
$W_{\bf M}$ have an 
exponentially small overlap with one another in the high-dimensional 
configuration space, in which case only the term ${\bf M}=0$ survives in the
 sum on the r.h.s. of Eq. \ref{omega_I_many_body_c}. Therefore we conclude that
for an insulator in any number of dimensions we have 
$\min [\Omega^{({\rm W})}] \doteq \left< {\Delta {\bf X}}^2 \right>$,
where for any finite size the corrections due to the exponentially small 
overlaps make the l.h.s. slightly larger than the r.h.s. Thus Kohn's functions
$\Psi_{\bf M}$ are the maximally-localized ``many-body Wannier functions'' 
$W_{\bf M}$, since as $V \rightarrow \infty$
the width $\left< {\Delta {\bf X}}^2 \right>$ of the gauge-invariant
distribution $p({\bf X})$ obtained from $\Psi_{\bf M}$ (Eq. \ref{p_X_bulk}) 
becomes the minimum of the width $\Omega^{({\rm W})}$
of the gauge-dependent distribution $p_{\rm W}({\bf X})$ obtained from
$W_{\bf M}$ (Eq. \ref{p_W}).

\section{
the quantum of polarization}
\label{quantum}

Here we will discuss the quantum of polarization\cite{ksv,om94} in terms of
Kohn's theory of localization\cite{kohn,kohn68}  
[A related discussion, regarding the closely related phenomenom of quantized 
charge transport in insulators, can be found in Ref. \cite{niu91}.]
Let us consider a periodic system of volume $L^3$, which for simplicity we
take to be cubic. If the external potential acting on 
the electrons is changed adiabatically along an insulating closed path 
parametrized by $\lambda$
$( \hat{H}^{(\lambda=1)} = \hat{H}^{(\lambda=0)} )$, then, since 
the hamiltonian comes back to itself, the net effect, as far as the 
wavefunction is concerned, has to be either (i) each $\Psi_{\bf M}$ returns to
itself (modulo a global phase, the same for all ${\bf M}$), or
(ii) apart from a global phase, there is a rigid translation of all the 
$\Psi_{\bf M}$ in configuration space by the same amount, which can be
described by a uniform shift of their indices: 
${\bf M} \rightarrow {\bf M}+(n_1,n_2,n_3)$.

As far as charge transport is concerned, the important observation is that, 
since the system remains insulating throughout the path, the
regions ${\cal R}_{\bf M}$ remain disconnected, so that no charge can 
flow between them during the adiabatic motion. The resulting
integrated current flowing through the system
during the cycle (which, according to Eqs. \ref{delta_P} and \ref{current1},
measures the change in polarization) can then be 
inferred from Figs. 2 and 3, which 
show that the
regions ${\cal R}_{\bf M}$ are labelled by the center of mass of the electrons,
and that one can go from one region to the next by moving any one electron
across the length $L$ of periodic the system.  
The smallest non-zero change in the average polarization along the $i$-th
direction is given by the smallest non-zero shift in the 
distribution $p(X_i)$ ($n_i=\pm 1$), and
is seen to equal $|q_e|L/V = |q_e|/L^2$, which is the quantum\cite{ksv,om94}.

It should be noted that 
strictly speaking the exact quantization of charge transport for an insulating
system with periodic boundary conditons 
only occurs in the thermodynamic limit\cite{thouless83,niu84}. In fact, the 
quantization was established for a finite size only after averaging over all 
twisted boundary conditions; for a finite size and periodic boundary 
conditions, there are exponentially small corrections\cite{niu84}. This is consistent with Kohn's picture that for any finite $V$ the regions 
${\cal R}_{\bf M}$ are not completely disconnected but have an exponentially 
small overlap, which allows for a correspondingly small charge flow between 
neighboring regions, thus destroying the exact quantization.

\section{the case of fractional filling}\label{frac_fill}

The formulas given in the text need to be generalized in order to deal with
correlated systems
which have in the ground state a non-integer number of 
electrons per primitive cell, and yet are insulating.
This can happen in many 1D 
models\cite{aligia99b,giamarchi97}, but it appears that in higher dimensions 
the Mott
transition to an insulating state is usually accompanied by a breaking of the
symmetry (e.g., a charge-density wave) which  restores integer 
filling\cite{shankar90}. Therefore we shall restrict the ensuing analysis to 
1D, as done in Refs. \cite{aligia99a,aligia99b}.

Aligia\cite{aligia99a} has shown that in cases where there is a fractional
number of electrons per primitive cell, the limits of integration over $k$ in 
the Berry phase formula for the polarization difference 
(Eq. \ref{delta_P_om94}) given in Ref. \cite{om94} need to be modified, from 
which it follows that the ``single-point'' formulas for the polarization
and localization derived in Refs. \cite{resta98,resta99} also need to be 
changed. For the purposes of the present paper it is straightforward to modify
the integral over $k$ in the definition of $\ln C(\alpha)$, and from that 
derive the required modifications to the discretized numerical formulas.

Let us consider a 1D system with a simulation cell of size $L$ and $n/l$ 
electrons per unit cell, where $n/l$ is an irreducible fraction. Following 
Ref. \cite{aligia99a}, we modify the cumulant generating function, 
Eq. \ref{bulk_M}, as follows:

\begin{equation}
\label{frac_cgf}
\ln C(\alpha) = \frac{1}{l} \int_0^{\frac{2 \pi l}{L}} dk 
\ln {\cal C}(k,\alpha).
\end{equation} 

\noindent In order to obtain the discretized formulas, we just need to retrace
the steps taken in Sections \ref{discrete_avg} and \ref{discrete_fix}, with the
above modification. Discretizing the interval
$0 \leq k \leq 2 \pi l/L$ into a uniform row of $J+1$ points $k_{\gamma}$, 
we find that Eq. \ref{polarization_discrete} changes to

\begin{equation}
\label{avg_discrete_frac}
\left< X \right> \simeq - \frac{L}{2 \pi l} {\rm Im} \ln \prod_{\gamma = 0}^{J-1}
\left< \Phi_{k_{\gamma}} \big| \Phi_{k_{\gamma+1}} \right>,
\end{equation}

\noindent which agrees with the result quoted in Ref. \cite{aligia99a}.
Similarly, Eq. \ref{variance_scaled} becomes:

\begin{equation}
\label{variance_scaled_frac}
\left< {\Delta X}^2 \right> \simeq - \frac{JL^2}{{(2\pi l)}^2}
\ln {\left| \prod_{\gamma=0}^{J-1} 
\left< \Phi_{k_{\gamma}} \big| \Phi_{k_{\gamma+1}} \right> 
\right|}^2. 
\end{equation}

\noindent As for the modified ``single-point'' formulas\cite{aligia99b}, it is 
straightforward to verify, applying the approach of Section \ref{discrete_fix}
to the previous two equations, that they are the same as
Eqs. \ref{avg_dscrt_sngl_pt} and \ref{variance_sngl_pt}, except for the 
substitution $\tilde{L} \rightarrow \tilde{L} / l$.

\break

FIGURE 1: Schematic representation of the regions
${\cal R}_{{\bf m}_1...{\bf m}_N}$ in the many-electron configuration space 
where the wavefunction of a finite system obeying periodic boundary conditions
is localized. The system has linear dimensions $\sim a$, and the periodic
boundary conditions are over a length $L >> a$, so that the regions
${\cal R}_{{\bf m}_1...{\bf m}_N}$ are essentially non-overlapping.\\ \\

FIGURE 2: Schematic representation of the essentially disconnected regions
${\cal R}_{\bf M}$ in the many-electron configuration space where the
wavefunction of an extended insulator is localized 
(adapted from Ref. \cite{kohn68}).
For $d$ real-space dimensions ${\bf M}=(M_1,...,M_d)$.
Shown is the case of two electrons and $d=1$, for which the
configuration space is $(x_1,x_2)$. The system is composed of 
possibly strongly overlapping units in real
space (e.g., a covalent insulator) and yet, because it is insulating, 
in configuration
space the wavefunction $\Psi$ breaks up into a sum of partial functions 
$\Psi_{\bf M}$, each localized in a region ${\cal R}_{\bf M}$, which have
an exponentially small overlap with one another, if the system is large.
\\ \\

FIGURE 3: Localized distribution $p(X_i)$ along the $i$-th direction of $N$ 
times the electronic center of mass (${\bf X} = \sum_{j=1}^N {\bf x}_j$) for a
$d$-dimensional insulator with $N$ electrons in a periodic volume $L^d$
(based on Fig. 6 of Ref. \cite{kohn}).
Although each individual electron coordinate ${\bf x}_j$, as well as the
electronic charge density (not shown),
are spread over the whole system, because the system is insulating a localized 
distribution $p(X_i)$ of width $\sqrt{N} \xi_i$ can be uniquely defined in 
terms of the partial wavefunction $\Psi_{\bf M}$ in a {\it single} 
disconnected region ${\cal R}_{\bf M}$ in the $dN$-dimensional 
configuration space (Fig. 2). Choosing a different region 
${\cal R}_{{\bf M}^{'}}$ simply shifts the center of the distribution, if 
$M^{'}_i \not= M_i$. The solid lines correspond to $d=1$, for which
the peaks coming from different ${\cal R}_{\bf M}$
do not overlap with one another for large $L$, whereas for $d=3$
(dotted lines) they overlap strongly  
(but the regions ${\cal R}_{\bf M}$ are still essentially 
disconnected, for large $L$.) \\ \\ \\

\begin{table}
\caption{Comparison between the formulas for the Drude weight and for the 
localization length, their relation to the optical conductivity, and their 
asymptotic values in the thermodynamic limit for insulators and conductors at 
$T=0$.}
\begin{tabular}{| c | c | c |}
& & \\
& Drude weight & Localization length \\
& & \\
\hline
& & \\
Formula in terms of & & \\
twisted boundary conditions & 
$D_i = \frac{1}{2V} {\left. \frac{\partial^2 E({\bf k})}{\partial k_i^2}
\right|}_{{\bf k}=0}$
& $\xi_i^2(N) = - \frac{1}{N} {\left. \frac{\partial^2 
\ln C({\boldsymbol \alpha})}{\partial \alpha_i^2} 
\right|}_{{\boldsymbol \alpha}=0}$ \\
& & \\
\hline 
& & \\
Relation to conductivity & 
$D_i = -\frac{1}{2} \lim_{\omega \rightarrow 0} \omega {\rm Im}
\sigma_{ii}(\omega)$ & 
$\xi_i^2(N) = \frac{\hbar}{\pi q_e^2 n_0} \int_0^{\infty} \frac{d \omega}
{\omega} {\rm Re} \sigma_{ii}(\omega)$ \\
& & \\
\hline
& & \\
Asymptotic value & & \\ ($N,V \rightarrow \infty$)  & & \\
& & \\
Insulators & Zero & Finite \\
Non-ideal conductors & Zero & Infinite \\
Ideal conductors & Finite & Infinite \\
& & \\
\end{tabular}
\end{table}


\begin{references}

\bibitem{landau_em} L. D. Landau and E. M. Lifshitz, 
{\em Electrodynamics of Continuous Media}, Second Edition
(Pergamon, New York, 1984).

\bibitem{jackson} J. D. Jackson, {\em Classical Electrodynamics}, Second 
Edition (Wiley, New York, 1975).

\bibitem{am} N. W. Ashcroft and N. D. Mermin, {\em Solid State Physics}
(Saunders, New York, 1976).

\bibitem{szigeti} B. Szigeti, Trans. Faraday Soc. {\bf 45}, 155 (1949).

\bibitem{harrison80} W. A. Harrison, {\em Electronic Structure and the
Properties of Solids} (Freeman, New York, 1980).

\bibitem{jennison} D. R. Jennison and A. B. Kunz, Phys. Rev. B {\bf 13}, 5597
(1976). 

\bibitem{pendry} J. B. Pendry and C. H. Hodges, J. Phys. C {\bf 17}, 1269 
(1984).

\bibitem{kohn} W. Kohn, Phys. Rev. {\bf 133}, A171 (1964).

\bibitem{kohn68}
W. Kohn, in {\em Many-Body Physics}, edited by C. DeWitt and R. Balian
(Gordon and Breach, New York, 1968), p. 351.

\bibitem{vinay} V. Ambegaokar and W. Kohn, Phys. Rev. {\bf 117}, 423 (1960).

\bibitem{pick} R. Pick, M. H. Cohen, and R. M. Martin, Phys. Rev. B {\bf 1},
910 (1970).

\bibitem{ksv} R. D. King-Smith and D. Vanderbilt, Phys. Rev. B {\bf 47}, 1651
(1993).

\bibitem{vks} D. Vanderbilt and R. D. King-Smith, Phys. Rev. B {\bf 48}, 4442
(1993).

\bibitem{om94} G. Ortiz and R. M. Martin, Phys. Rev. B {\bf 49}, 14 202 
(1994).

\bibitem{resta94} R. Resta, Rev. Mod. Phys. {\bf 66}, 899 (1994).

\bibitem{resta_lect} R. Resta, {\em Berry Phases in Electronic Wavefunctions},
Trosi\`eme Cycle Lecture Notes (\'Ecole Polytechnique F\'ed\'erale, Lausanne,
Switzerland, 1996); also available online at 
http://ale2ts.ts.infn.it:6163/$\sim$resta/publ/notes\_trois.ps.gz

\bibitem{resta98} R. Resta, Phys. Rev. Lett. {\bf 80}, 1800 (1998).

\bibitem{rmm74} R. M. Martin, Phys. Rev. B {\bf 9}, 1998 (1974).

\bibitem{hirst} L. L. Hirst, Rev. Mod. Phys. {\bf 69}, 607 (1997).

\bibitem{rmm72} R. M. Martin, Phys. Rev. B {\bf 5}, 1607 (1972);
{\bf 6}, 4874 (1972).

\bibitem{resta92} R. Resta, Ferroelectrics {\bf 136}, 51 (1992).

\bibitem{vogl78} P. Vogl, J. Phys. C {\bf 11}, 251 (1978).

\bibitem{blount} E. I. Blount, Solid State Phys. {\bf 13}, 305 (1962).

\bibitem{thouless83} D. J. Thouless, Phys. Rev. B {\bf 27}, 6083 (1983).

\bibitem{niu84} Q. Niu and D. J. Thouless, J. Phys. A {\bf 17}, 2453 (1984).

\bibitem{wannier37} G. H. Wannier, Phys. Rev. {\bf 52}, 191 (1937).

\bibitem{foot_one_point} Such an interpretation assumes that there exists
a centrosymmetric structure which is connected to the structure of 
interest by an insulating path in $\lambda$-space, and that a vanishing bulk 
polarization can be assigned to the centrosymmetric structure, on the basis of 
symmetry.

\bibitem{aligia99a} A. A. Aligia, Europhys. Lett. {\bf 45}, 411 (1999).

\bibitem{kudinov} E. K. Kudinov, Fiz. Tverd. Tela {\bf 33}, 2306 (1991)
[Sov. Phys. Solid State {\bf 33}, 1299 (1991)].

\bibitem{callen} H. B. Callen and T. A. Welton, Phys. Rev. {\bf 83}, 34 (1951).

\bibitem{callen52} H. B. Callen and M. L. Barasch, Phys. Rev. {\bf 88}, 1382
(1952).

\bibitem{landau} L. D. Landau and E. M. Lifshitz, {\em Statistical Physics}
(Pergamon, New York, 1980).

\bibitem{marzari97} N. Marzari and D. Vanderbilt, Phys. Rev. B {\bf 56},
12 847 (1997).

\bibitem{resta99} R. Resta and S. Sorella, Phys. Rev. Lett. {\bf 82}, 
370 (1999).

\bibitem{silvestrelli} See also P. L. Silvestrelli, N. Marzari,
D. Vanderbilt, and M. Parrinello, Solid State Commun. {\bf 107}, 7 (1998), 
where a similar functional was proposed in an independent-electron framework as
a measure of the spread of the Wannier functions.

\bibitem{aligia99b} A. A. Aligia and G. Ortiz, Phys. Rev. Lett. {\bf 82},
2560 (1999).

\bibitem{kendall} M. G. Kendall and A. Stuart, {\em The Advanced Theory of
Statistics}, Volume 1, Third Edition (Hafner, New York, 1969).

\bibitem{kubo62} R. Kubo, J. Phys. Soc. Japan {\bf 17}, 1100 (1962).

\bibitem{fulde} P. Fulde, {\em Electron Correlations in Molecules and Solids}
(Springer, Berlin, 1995).

\bibitem{foot_average} We will distinguish between two somewhat different 
usages of the notation $\left< ... \right>$: $\left< {\bf X} \right>$ denotes
the average of the {\it distribution} $p({\bf X})$, whereas 
$\left< \hat{\bf X} \right>$ denotes the quantum expectation value of
the {\it operator} $\hat{\bf X}$.

\bibitem{foot_dotproduct} The second form of Eq. \ref{precursor_mgf} will be
the most useful in what follows. We note that it is the many-body analogue of a
quantity introduced in the single-particle context in Ref. \cite{marzari97}
[see Eq. 25 of Ref. \cite{marzari97}].

\bibitem{foot_quantum} The first moment (average bulk polarization) is
actually somewhat special in this respect, since even the correct expression is
gauge-invariant only modulo a quantum, as discussed in 
Section \ref{recent_develop}. Hence what is meant here by lack of 
gauge-invariance is gauge-dependence apart from the quantum. 

\bibitem{weinreich} G. Weinreich, {\em Solids: Elementary Theory for Advanced
Students} (Wiley, New York, 1965).

\bibitem{berry89} M. V. Berry, in {\em Geometric Phases in Physics}, edited
By A. Shapere and F. Wilczek (World Scientific, Singapore, 1989), p. 7.

\bibitem{provost}  J. P Provost and G. Vallee, Commun. Math. Phys. {\bf 76},
289 (1980).

\bibitem{zak} J. Zak, Solid State Phys. {\bf 27}, 1 (1972); 
J. Zak, Phys. Rev. B {\bf 20}, 2228 (1979).

\bibitem{scalapino} D. J. Scalapino, S. R. White, and S. Zhang, Phys. Rev. B
{\bf 47}, 7995 (1993), and references therein.

\bibitem{foot_divergence} For a finite $V$ the levels of any 
system are discrete, and therefore absorption becomes impossible as soon as 
$\omega$ becomes less than the energy spacing between the levels of the
system\cite{kohn68}, so that the fluctuations are in general finite even for a 
conductor, at any given ${\bf k}$. 
It is also clear that the crossing of low-lying levels at a finite number of
${\bf k}$-points in a conductor
at finite $V$\cite{scalapino} does not affect our conclusion regarding the
divergence of Eq. \ref{fdt_bulk} as $V \rightarrow \infty$.

\bibitem{stafford91} C. A. Stafford, A. J. Millis, and B. S. Shastry,
Phys. Rev. B {\bf 43}, 13 660 (1991).

\bibitem{bassani} F. Bassani and G. Pastori Parravicini, {\em Electronic States
and Optical Transitions in Solids} (Pergamon, Oxford, 1975), Chap. 5.

\bibitem{kivelson} S. Kivelson, Phys. Rev. B {\bf 26}, 4269 (1982).

\bibitem{foot_continuum} $N_c$ equals the number of cells in the periodic 
crystal; when that number goes to infinity the sum is replaced by an integral:
$(1/N_c) \sum_{\bf k} \rightarrow v/(2\pi)^3 \int d{\bf k}$ (in the 
correlated case $v$ is replaced by $V$).

\bibitem{foot_overlap} According to Eq. \ref{phi_wf}
$\Psi_{{\bf k}=0} \propto \sum_{\bf M} W_{\bf M}$. Since $\Psi_{{\bf k}=0}$ is
gauge-invariant apart from a global phase, if the $W_{\bf M}$ were 
non-overlapping, they would also have to be gauge-invariant, apart from the 
same global phase. 

\bibitem{foot_difference_kohn} In Appendix 1 of Ref. \cite{kohn} a
definition for the center-of-mass distribution of a periodic system was 
proposed, which amounts to replacing the $\delta$-function in
Eq. \ref{X_dist} by a suitably defined ``periodic $\delta$-function''. For a 1D
insulator such definition yields the 
same distribution (for large $L$) as the characteristic function $C(\alpha)$. 
However, it does not give an individual peak, since the resulting
distribution is periodic by construction; as a consequence, in 3D it
yields a delocalized center of mass even for insulators, because 
the contributions from different $\Psi_{\bf M}$
overlap with one another (this is the ``strange 
situation'' alluded to in Appendix 1 of Ref. \cite{kohn}); our definition in 
terms of the characteristic function, on the other hand, generates one 
localized distribution $p({\bf X})$ from each
$\Psi_{\bf M}$ separately (see overlapping dotted curves in Fig. 3), 
which retains the desired physical interpretation in any number of dimensions.

\bibitem{foot_log} Notice that Eq. \ref{first_moment_scaled} can be
rewritten without the logarithm:
${\left< S^l \right>} = (\imath/2 \pi) \int d{\bf q} {\left.
\partial_{t_l}{\cal C}({\bf q},{\bf t}) \right|}_{{\bf t}=0}$. However, if we
discretize the derivative in the integrand in the straightforward way,
the resulting expression will no longer be gauge-invariant.

\bibitem{foot_even_derivatives} However, it is easy to see that
$- {\left. \partial^2_{\alpha_i \alpha_j} 
{\rm Re} \ln {\cal C}({\bf k},{\boldsymbol \alpha})
\right|}_{{\boldsymbol \alpha}=0} = G_{ij}({\bf k})$, so that taking the real
part of Eq. \ref{second_der_discretized} yields a gauge-invariant 
approximation to the metric tensor in scaled coordinates. 

\bibitem{kohn96} W. Kohn, Phys. Rev. Lett. {\bf 76}, 3168 (1996).

\bibitem{mo97} R. M. Martin and G. Ortiz, Phys. Rev. Lett. {\bf 78},
2028 (1997). 

\bibitem{foot_composite} Here all the filled bands are being treated as a 
{\it composite band} (in the terminology of Ref. \cite{marzari97}), to
which corresponds a single $\Omega_{\rm I}$; in general different
$\Omega_{\rm I}$'s can be ascribed to each separate group of composite 
bands\cite{marzari97}, but then the connection to the polarization
fluctuations, which involve all the valence electrons, appears to be lost.

\bibitem{foot_kivelson} In Ref. \cite{kivelson} the Wannier functions were
obtained as the eigenfunctions of the position operator projected into a given
band. As shown in Ref. \cite{marzari97}, in 1D this choice minimizes the 
quadratic spread. Notice also that Kivelson considered disordered insulators;
in that case $E_{\rm g}$ is the absolute gap.


\bibitem{niu91} Q. Niu, Mod. Phys. Lett. B {\bf 5}, 923 (1991).

\bibitem{giamarchi97} T. Giamarchi, Physica (Amsterdam) {\bf 230B-232B}, 975
(1997).

\bibitem{shankar90} D. H. Lee and R. Shankar, Phys. Rev. Lett. {\bf 65},
1490 (1990).

\end{references}
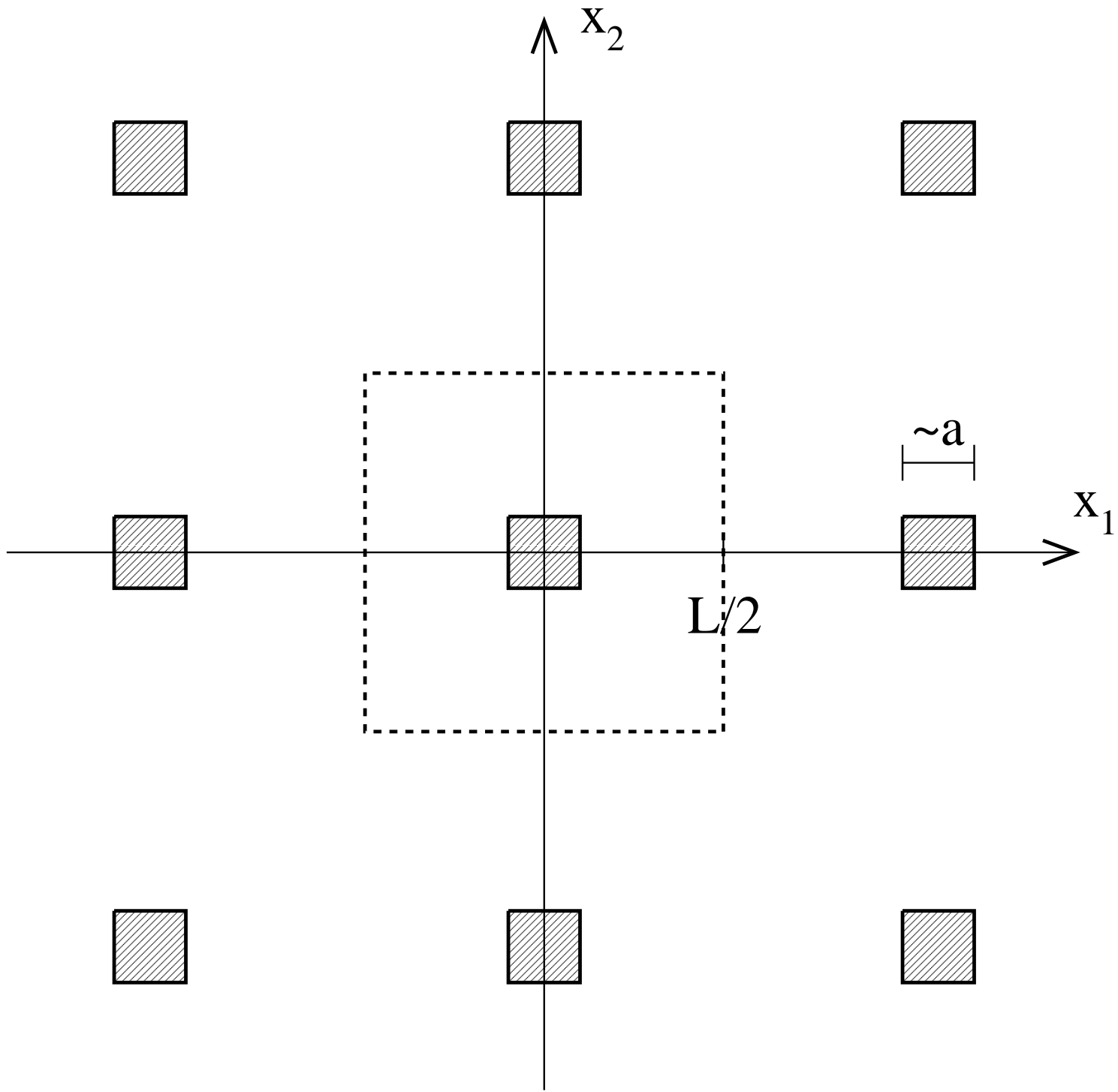
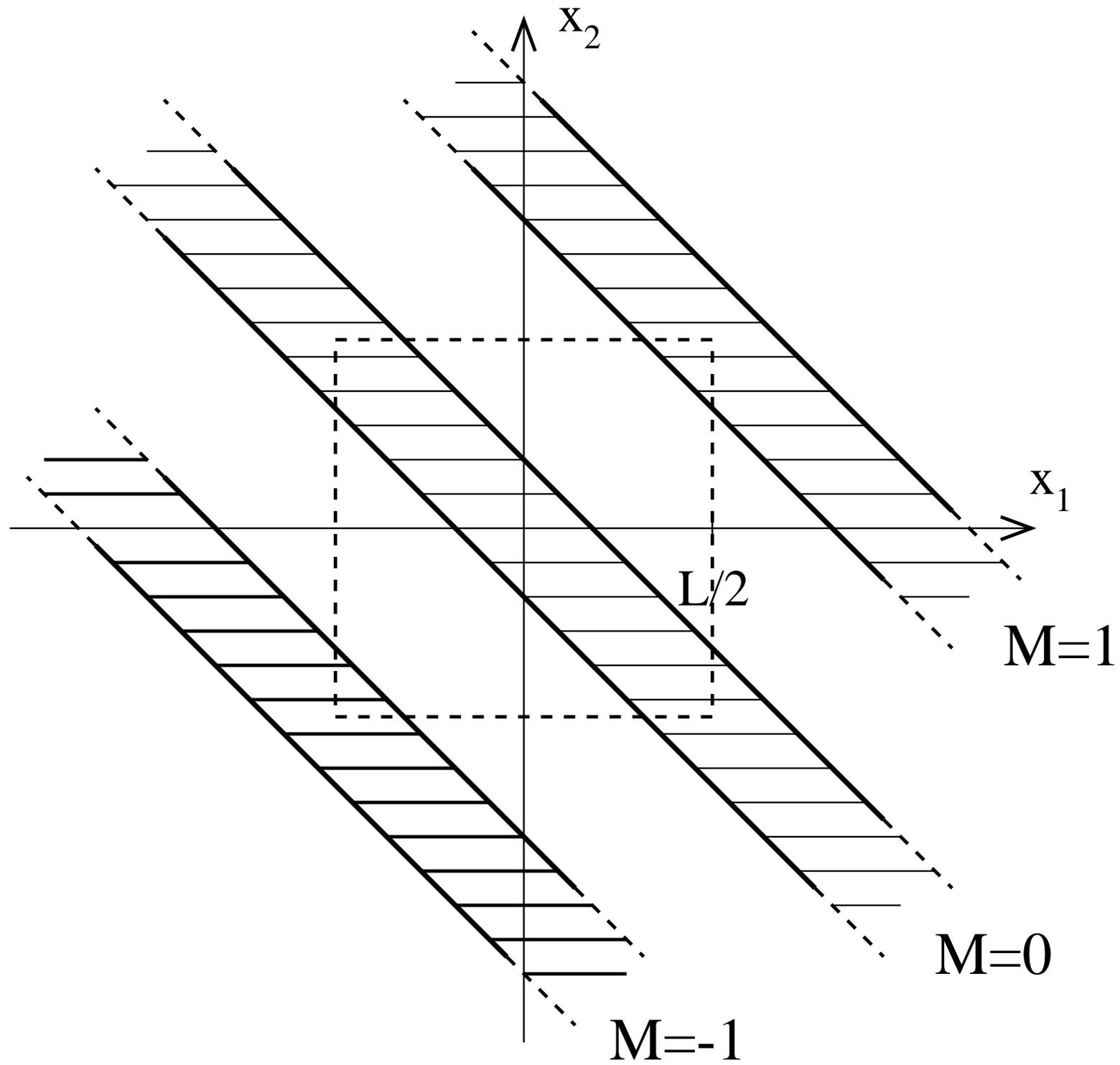
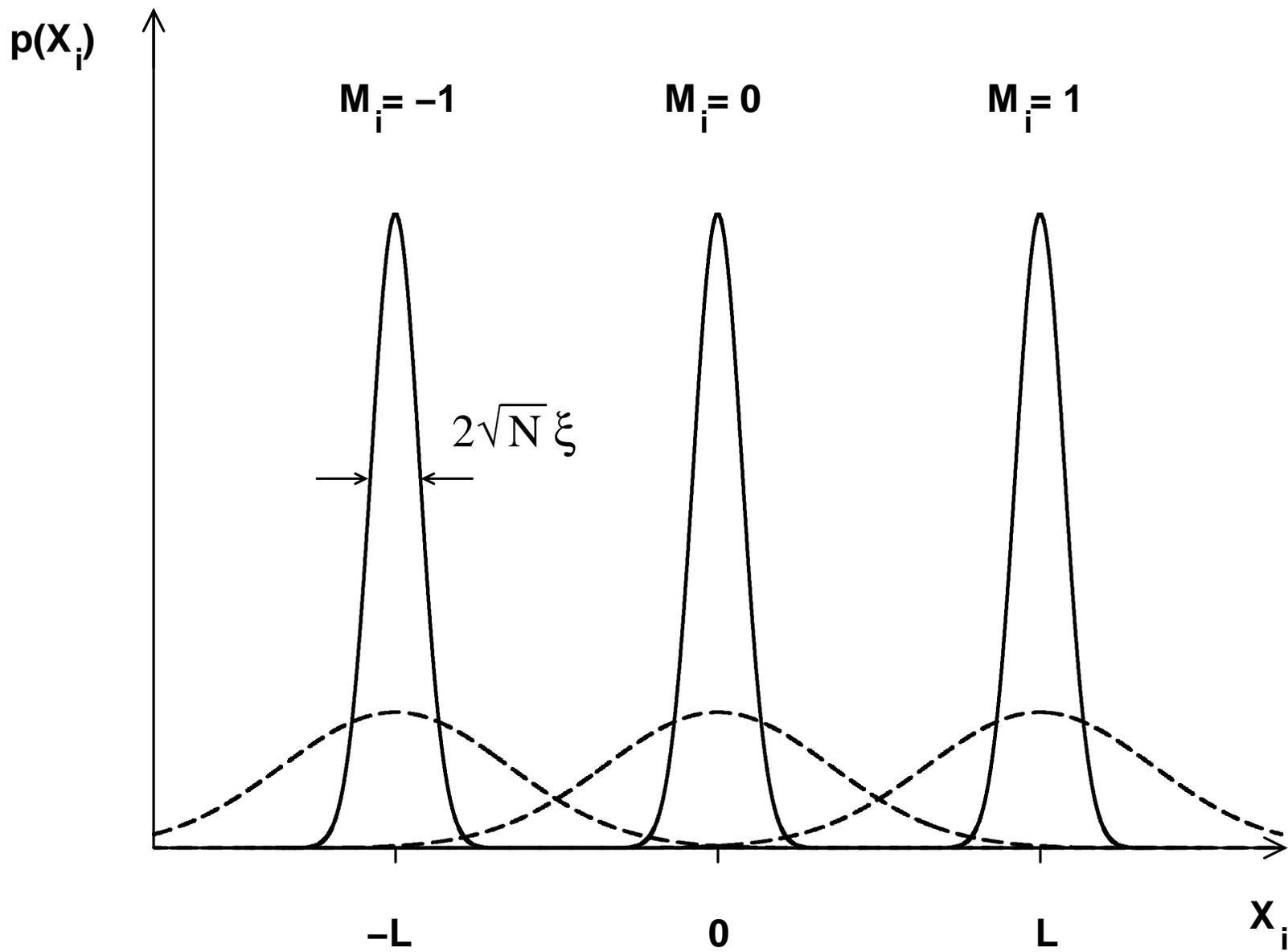
\end{document}